\documentclass[a4paper,11pt]{article}
\pdfoutput=1  
\usepackage{slashed}
\usepackage{jcappub} 
\usepackage[T1]{fontenc}  
\usepackage{graphicx}
\usepackage{amsmath}
\usepackage{textcomp}
\usepackage{gensymb}
\usepackage{subcaption}
\def\red{\color{black}}

\usepackage{float}
\usepackage{mwe}

\title{\boldmath Tomographic analyses of the CMB lensing and galaxy clustering to probe the linear structure growth}

\author[a,b,1]{Gabriela A. Marques,\note{Corresponding author.}}
\author[a]{Armando Bernui}

\affiliation[a]{Observat\'orio Nacional, Rua General Jos\'e Cristino 77, S\~ao Crist\'ov\~ao, 20921-400 Rio de Janeiro, RJ, Brazil}
\affiliation[b]{Department of Physics, Florida State University, Tallahassee, Florida 32306, USA}

\emailAdd{gmarques@fsu.edu}
\emailAdd{bernui@on.br}

\abstract{
In a tomographic approach, we measure the cross-correlation between the CMB lensing reconstructed from the Planck satellite and the galaxies of the photometric redshift catalogue based on the combination of the South Galactic Cap u-band Sky Survey (SCUSS), Sloan Digital Sky Survey (SDSS), and Wide-field Infrared Survey Explorer (WISE) data. We perform the analyses considering six redshift bins spanning the range of $0.1 <z<0.7$. From the estimates of the galaxy-galaxy and galaxy-CMB lensing power spectrum, we derive the galaxy bias and the amplitude of the cross-correlation for each redshift bin. We have finally applied these tomographic measurements to estimate the linear structure growth using the bias-independent $\hat{D}_{G}$ estimator introduced by \cite{giannantonio2016cmb}. We find that the amplitude of the structure growth with respect to the fiducial cosmology is $A_{D}=1.16\pm0.13$, closely consistent with the predictions of the $\Lambda$CDM model ($A_{D}^{\Lambda \mbox{\footnotesize CDM}}=1$). We perform several tests for consistency of our results, finding no significant evidence for systematic effects.

}

\begin{document}
\maketitle 
\flushbottom
\section{Introduction}
\label{sec:intro}
 Progress in the sensitivity of astronomical photometric surveys dedicated to the study the large-scale structure (LSS) has been providing valuable information about the features of the Universe at several scales and redshifts \citep{dark2005dark,richards2008efficient,bilicki2013two,bilicki2016wise,beck2016photometric}. The prospects of using LSS data to constrain cosmology are very promising. Several upcoming astronomical surveys will produce extensive photometric data covering a wide area of the sky such as the Large Synoptic Survey Telescope (LSST) \citep{abellpa} and the Wide-Field Infrared Survey Telescope (WFIRST). On the other hand, the cosmic microwave background radiation (CMB) allow us to test the primordial characteristics of the Universe \citep{bernui2006temperature,novaes2014searching}. However, before reaching us, the CMB photons are affected by inhomogeneities along their path producing a range of secondary effects, beyond the primary CMB temperature fluctuations at the last scattering surface~\citep{Sachs:1967er, Sunyaev:1980vz, Rees:1968zza}. One of these secondary effects is the gravitational deflection of the CMB photons by the mass distribution along their path, namely weak gravitational lensing.\ 
 
The CMB lensing has been investigated by several methods and experiments in the past~ e.g, \citep{hirata2004cross,smith2007d,das2011detection,van2012measurement,das2014atacama,2014PhRvL.113b1301A,2017PhRvD..95l3529S}. 
Recently, through observations of the Planck satellite, it was possible not only to detect the lensing effect with high statistical significance but also to robustly reconstruct the lensing potential map in almost full-sky \citep{ade2014planck,ade2016xv,lensing2018planck}. Such a reconstructed map contains unique information  of the LSS since it is related to the integral of the photon deflections from us until the last scattering surface.
 
Although the CMB lensing signal covers a broad redshift range, from local to high redshifts, it is not possible to obtain the evolution of the LSS along the line of sight using only the CMB lensing data. 
However, the cross-correlation technique with another tracer of matter provides additional astrophysical and cosmological information. Several galaxy catalogs, such as those from the Wide Field Survey Infrared Explorer (WISE) and derived catalogues \citep{ade2014planck,krolewski2019unwise}, NRAO VLA Sky Survey (NVSS) ~\citep{hirata2008co}, Canada-France-Hawaii Telescope (CFHT)~\citep{omori2015cross}, Sloan Digital Sky Survey (SDSS)~\citep{singh2016cross,giusarma2018scale,singh2018probing},   2MASS~\citep{Bianchini2018,peacock2018wide}, Dark Energy Survey (DES)~\citep{2016MNRAS.461.4099B,giannantonio2016cmb,omori2018dark} and WISExSuperCOSMOS \citep{peacock2018wide,raghunathan2018imprints} have already been cross-correlated with the CMB lensing potential. Also there have been significant detections of the cross-correlation between CMB lensing and galaxy lensing shear maps e.g.,~\citep{liu2015cross,DESlenslens,2018sdsslens, polabear}. In addition, analysis in a deeper Universe has been extended through the cross-correlation of CMB lensing with density tracers at high redshifts, e.g. quasars \cite{sherwin2012atacama,geach2013direct,dipompeo2014weighing} and sub-mm galaxies from Herschel H-ATLAS survey \citep{bianchini2015cross,2019arXiv190307046A}. Particularly, the galaxy auto-correlation and cross-correlation with the CMB lensing provides the opportunity to constrain the linear growth of the density fluctuations. By probing the evolution of perturbations over time it is possible to understand the mechanism that sources the late-time accelerated expansion of the Universe and shed light to distinguish a variety of gravity models~\cite{zhang2007probing,reyes2010confirmation,pullen2015probing}.\
Analysis of the Redshift-Space Distortions \citep{kaiser1987clustering} from spectroscopic surveys is a traditional way to measure the linear growth rate since it is commonly parameterized on large-scales by $f\sigma_{8}$, where $f \equiv d$ln$D/d$ln$a$ is the logarithmic derivative of the growth factor with respect to the scale factor $a$, and $\sigma_{8}$ is the linear matter variance in a spherical shell of radius 8 Mpc $h^{-1}$. However, this quantity can not be accurately measured for photometric surveys. Besides the photometric data be complementary with respect to spectroscopic studies, many photometric surveys are currently in operation or planned for the near future, with the obvious advantages of being lower-cost surveys and capable of mapping large areas of the sky with optimal observation strategies. Therefore, the $\hat{D}_{G}$ statistic introduced by \cite{giannantonio2016cmb} establishes an alternative to measure the linear growth for photometric redshift surveys. This estimator combines properly the auto and cross-correlation of the galaxy clustering and CMB lensing in such a way that it is bias-independent on linear scales.\

The aim of the present work is to constrain the linear growth factor $D$ in a tomographic approach and, therefore, to measure the evolution of the linear growth function. For that, we consider the $\hat{D}_{G}$ statistic using the CMB lensing map reconstructed by the $Planck$ team \cite{ade2016xv} with the galaxy overdensities from a multi-band photometric data released by \citep{gao2018photometric}, based on imaging from South Galactic Cap u-band Sky Survey (SCUSS), SDSS and WISE. The analyses are performed in six redshift bins spanning $0.1 < z < 0.7$, being complementary to the linear growth measures previously found for others photometric catalogs ~\citep{giannantonio2016cmb, Bianchini2018,peacock2018wide,omori2018dark}. Although the $\hat{D}_{G}$ estimator is galaxy bias independent, additionally, we use the measured galaxy-CMB lensing cross-correlations and galaxy auto-correlation to infer the correlation amplitude and the linear bias over the redshift bins.\ 

This paper is structured as follows : Firstly, we introduce the theoretical formalism in sec.~\ref{sec:theory}. In sec.~\ref{sec:data} we summarize the data used in the analysis. In sec.~\ref{sec:method} we describe the methodology. We then present the results and explore possible systematics and null tests in sec.~\ref{sec:results}, and our conclusions in sec.~\ref{sec:conclusion}.

\section{Background}
\label{sec:theory}
The gravitational lensing effect remaps the CMB temperature anisotropies by a angular gradient of the lensing potential, $\alpha(\mathbf{\hat{n}}) = \nabla \psi(\mathbf{\hat{n}})$, where $\nabla$ is the 2D gradient operator on the sphere and $\psi(\mathbf{\hat{n}})$ is the lensing potential. The Laplacian of the lensing potential is related to the convergence $\kappa(\mathbf{\hat{n}})$, which can be written as a function of the three-dimensional matter density contrast $\delta$ (see e.g.~\citep{ bartelmann2001weak}) 
\begin{equation}
    \kappa(\hat{n}) = \int_{0}^{\infty} dz W^{\kappa}(z) \delta(\chi(z)\mathbf{\hat{n}},z),
\end{equation}
where the CMB lensing kernel $W^{\kappa}$ is
\begin{equation}
    W^{\kappa}(z)= \frac{3 \Omega_{m}}{2c}\frac{H_{0}^2}{H(z)}(1+z) \chi(z)\frac{\chi_{*}-\chi(z)}{\chi_{*}}.
\end{equation}
Here we are considering a flat universe, $c$ is the speed of light, $H(z)$ is the Hubble parameter at redshift $z$, and $\Omega_{m}$ and $H_{0}$ are the present-day parameters of the matter density and Hubble, respectively. The comoving distances $\chi(z)$ and $\chi_{*}$ are set to the redshift $z$ and to the last scattering surface at $z_{*}\simeq 1090$, respectively. 
  
On the other hand, the galaxy overdensity $\delta_{g}$ from a galaxy catalogue with normalized redshift distribution $dn/dz$ also provides an estimate of the projected matter density contrast, given by
\begin{equation}
    \delta_{g}(\mathbf{\hat{n}}) = \int_{0}^{\infty} dz  W^{g}(z)\delta(\chi(z)\mathbf{\hat{n}},z),
\end{equation}
where the galaxy kernel $W^{g}$ for a linear, deterministic and scale-independent galaxy bias $b(z)$ \citep{fry1993biasing} is 
 \begin{equation}
    W^{g}(z) = b(z)\frac{dn}{dz}.
 \end{equation}
 Under the Limber approximation \citep{limber1953analysis}, the two-point statistics in the harmonic space of the galaxy-galaxy and galaxy-CMB lensing correlations become 
  \begin{equation}
\begin{aligned}
C_{\ell}^{gg} &= \int_{0}^{\infty}\frac{dz}{c}\frac{H(z)}{\chi^2(z)}[W^{g}(z)]^2 P\bigg(k=\frac{\ell+\frac{1}{2}}{\chi(z)},z \bigg), \\
C_{\ell}^{kg} &= \int_{0}^{\infty}\frac{dz}{c}\frac{H(z)}{\chi^2(z)}W^{\kappa}(z)W^{g}(z) P\bigg(k=\frac{\ell +\frac{1}{2}}{\chi(z)},z \bigg),
\end{aligned}
\label{eq:powerspe_th}
\end{equation}
where $P(k,z)$ is the matter power spectrum. The Limber approximation is quite accurate when $\ell$ is not too small ($\ell >10$) \citep{limber1953analysis}, which is the regime considered in this work. Moreover, is possible to rewrite the equations \ref{eq:powerspe_th} in terms of the linear growth function $D(z)$, since $P(k,z) = P(k,0)D^2(z)$. Therefore,
\begin{equation}
\begin{aligned}
C_{\ell}^{gg} \propto b^2(z)D^2(z), \\
C_{\ell}^{kg} \propto b(z)D^2(z).
\end{aligned}
\label{eq:rel}
\end{equation}
Thus, by properly combining the two quantities of the equation \ref{eq:rel}, it is possible to break the degeneracy between the galaxy bias and the linear growth through the estimator introduced by \cite{giannantonio2016cmb}:
\begin{equation}
\hat{D}_{G}\equiv\bigg\langle \frac{(C_{\ell}^{\kappa g})_{obs}}{(\slashed{C}_{\ell}^{\kappa g})_{th}}\sqrt{\frac{(\slashed{C}_{\ell}^{gg})_{th}}{(C_{\ell}^{gg})_{obs}}} \bigg\rangle_{\ell} .
\label{eq:dg}
\end{equation}
In the above equation, the slashed quantities, $\slashed{C}_{\ell}^{gg}$ and $\slashed{C}_{\ell}^{\kappa g}$ denote the theoretical functions evaluated at $z=0$ such that the growth function dependency is removed and $\hat{D}_{G}$ is normalized to $z=0$ ($\hat{D}_{G}(z=0) = 1$).

In order to obtain the theoretical predictions for the matter power spectrum $P(k,z)$, we use the public Boltzmann code CAMB\footnote{\url{https://camb.info/}} \citep{lewis2011camb} with the \texttt{Halofit} \citep{smith2003stable} extension to nonlinear evolution. Throughout the paper, we use the Planck 2018 cosmology \citep{planck2018params} derived by the \texttt{TT,TE,EE+lowE+lensing} data, with parameters $\{\Omega_{b}h^2, \Omega_{c}h^2, \Omega_{m}, \tau, n_{s}, A_{s},h\}$ = $\{0.0223,  0.1200,$ $0.3153,0.054, 0.964, 2.1\times 10^{-9}, 0.673\}$.
 
\section{Data}
\label{sec:data}
\subsection{The Galaxy catalogue}
In this study, we use the photometric redshift catalogue publicly released by \citep{gao2018photometric}. 
This catalog is based on multi-band data from three independent surveys: the South Galactic 
Cap u-band Sky Survey (SCUSS; \citep{zhou2016south2}), Sloan Digital Sky Survey (SDSS; 
\citep{york2000sloan}), and Wide-field Infrared Survey Explorer (WISE; \citep{wright2010wide}). Below, we briefly describe the properties of each survey and of the final catalog used in our analysis.\

The SCUSS is a u-band (354 nm) imaging survey using the 2.3m Bok telescope located on Kitt Peak, USA. The data products were released in 2015 containing calibrated single-epoch images, stacked images, photometric catalogs, and the star proper motions. The survey covers an area of approximately 5000 deg$^2$ of the South Galactic Cap and overlaps roughly $75\%$ of the area covered by the SDSS \citep{zou2016south}. The detailed information about the SCUSS and the data reduction can be found in \citep{zou2016south} and \citep{zou2015south}.\

The SDSS is a multi-spectral imaging and spectroscopic redshift survey, encompassing an area of about 14000 deg$^2$. The SDSS uses a wide-field camera that is made up of 30 CCDs. The survey is carried out imaging in five broad bands $u, g, r, i, z$, with limit-magnitude with $95\%$ completeness 22.0, 22.2, 22.2, 21.3 and 20.5 mag, respectively. 
The data have been released publicly in a series of roughly annual data releases. 
Specifically, the photometric data from the SDSS Data Release 10 (DR10) \citep{ahn2014tenth} was considered to obtain the galaxy catalogue used in this paper, as detailed in \citep{gao2018photometric}.

WISE is an infrared astronomical space telescope that scanned all-sky at 3.4, 4.6, 12 and 22 $\mu$m, known as W1, W2, W3, and W4, respectively. In September 2010, the frozen hydrogen cooling the telescope was depleted and the survey continued as NEOWISE, with the W1 and W2 bands. In order to match properly the official all-sky WISE catalogs with the SDSS data, is considered a technique to measure model magnitudes of the SDSS objects in new coadds of WISE images, called as \textit{forced photometry,} providing an extensive extragalactic catalogue of over 400 million sources \citep{lang2014unwise,lang2016wise}.
 
The catalogue we use has been built by combining the 7 photometric bands ranging from the near-ultraviolet to near-infrared. A local linear regression algorithm \citep{beck2016photometric} is adopted with a spectroscopic training set composed mostly of galaxies from the SDSS DR13 spectroscopy, in addition to several other surveys. The model magnitudes utilize the shape parameters from SDSS r-band and also the SDSS star/galaxy separation to characterize the source type. 
After correcting for galactic extinction~\citep{schlegel}, the final catalogue contains $\sim$ 23.1 million galaxies \footnote{Available for download from \href{http://batc.bao.ac.cn/~zouhu/doku.php?id=projects:photoz:start}{http://batc.bao.ac.cn/~zouhu/doku.php?id=projects:photoz:start}} with $\sim 99\%$ of the sources spanning the redshift interval of $z\leq 0.9$ \citep{gao2018photometric}. The use of a deeper SCUSS u-band and the multi-band information allowed the photo-z estimate more accurately and less biased than the SDSS photometric redshifts, with the average bias of $\overline{\Delta z_{norm}}= 2.28\times10^{-4}$ and standard deviation of $\sigma_{z}= 0.019$.\ 
Such a relatively deep catalog, with a remarkable galaxy number density, accurate photo-z, and considerable sky area, allows to use it to perform tests of the structure growth.

\begin{figure*}[!htbp]
    \centering
    \begin{subfigure}[b]{0.475\textwidth}
        \centering
        \includegraphics[width=\textwidth]{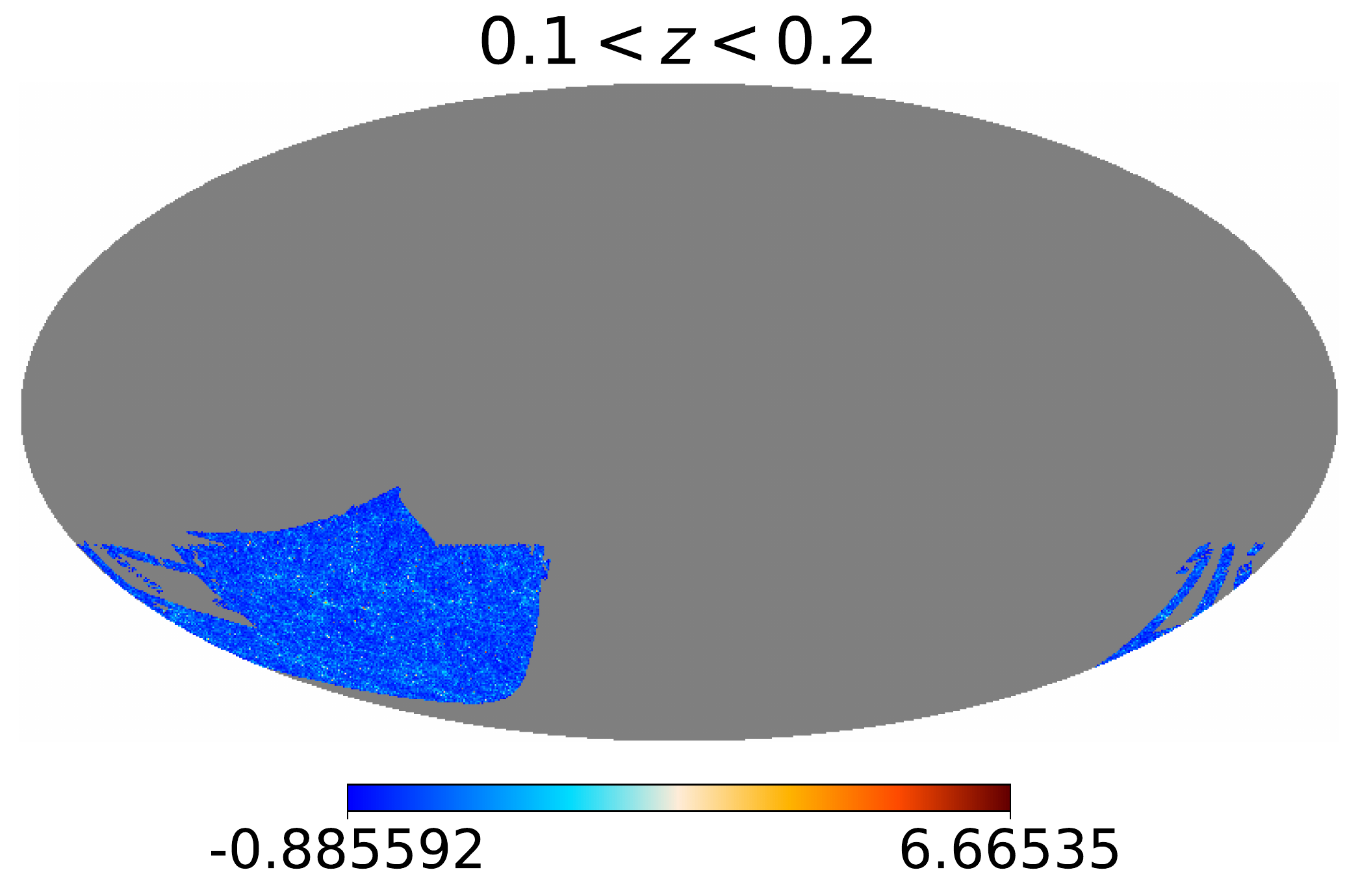}
    \end{subfigure}
    \begin{subfigure}[b]{0.475\textwidth}  
        \centering 
        \includegraphics[width=\textwidth]{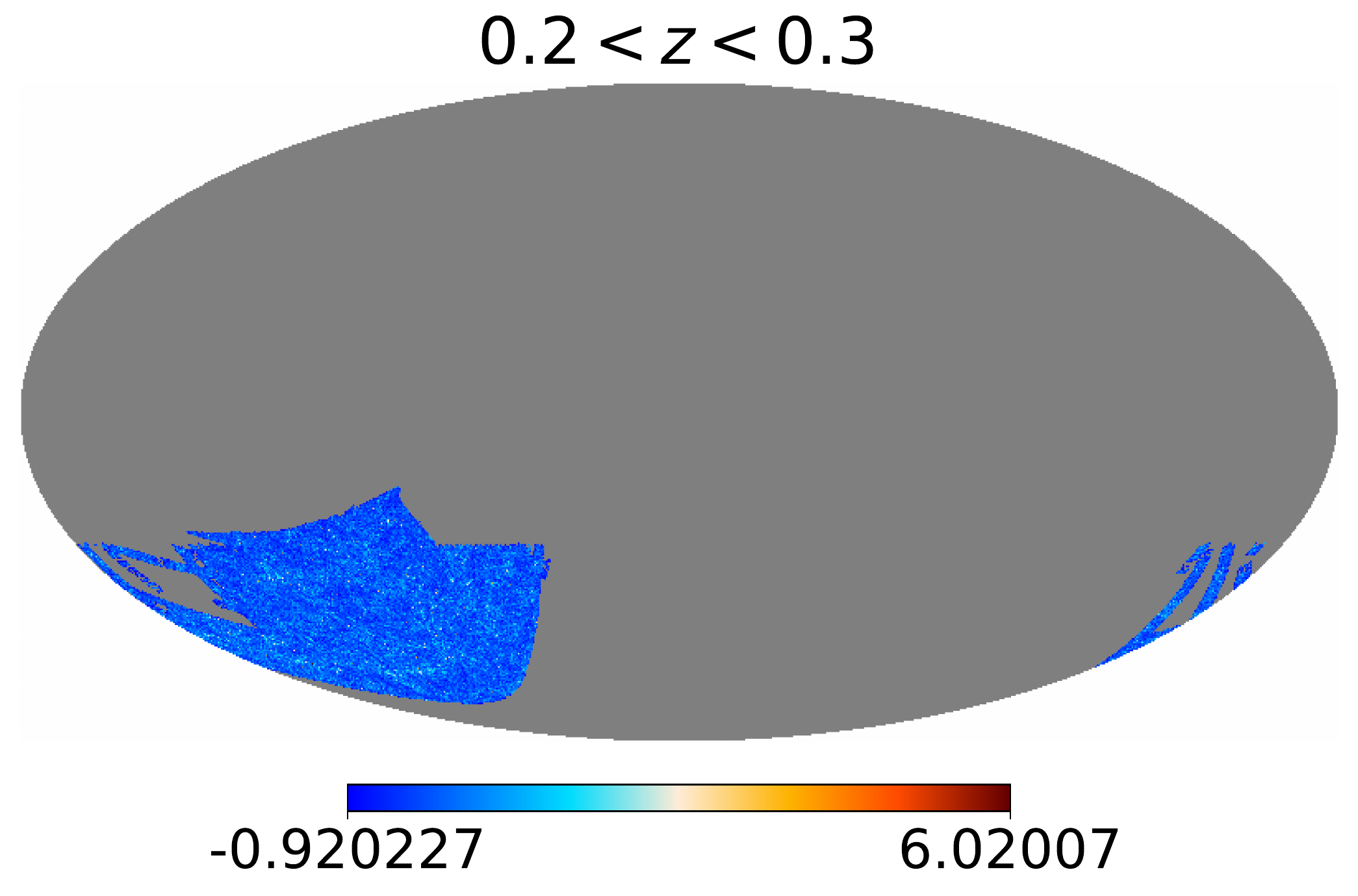}
    \end{subfigure}
    \begin{subfigure}[b]{0.475\textwidth}   
        \centering 
        \includegraphics[width=\textwidth]{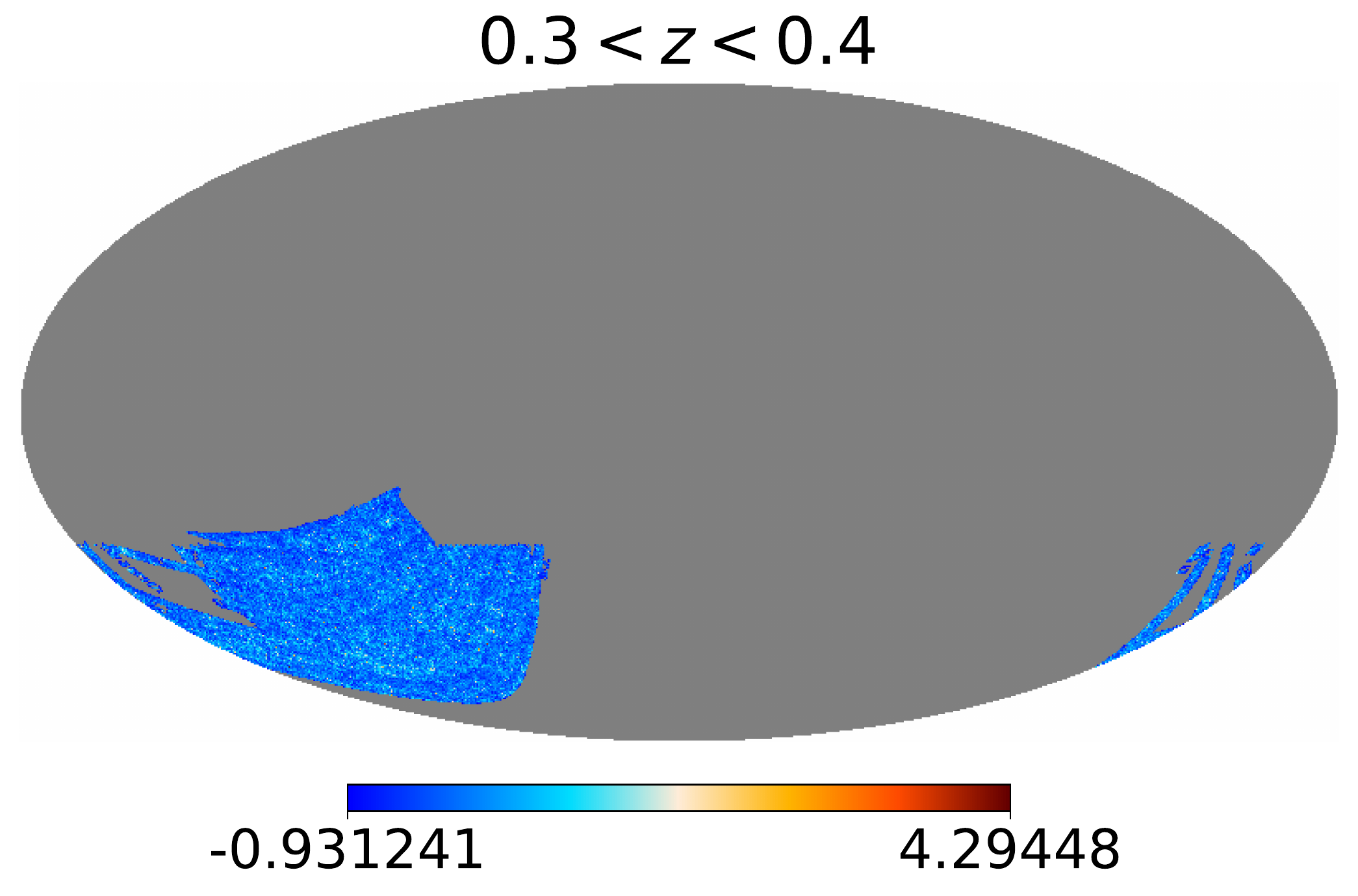}
    \end{subfigure}
    \begin{subfigure}[b]{0.475\textwidth}   
        \centering 
        \includegraphics[width=\textwidth]{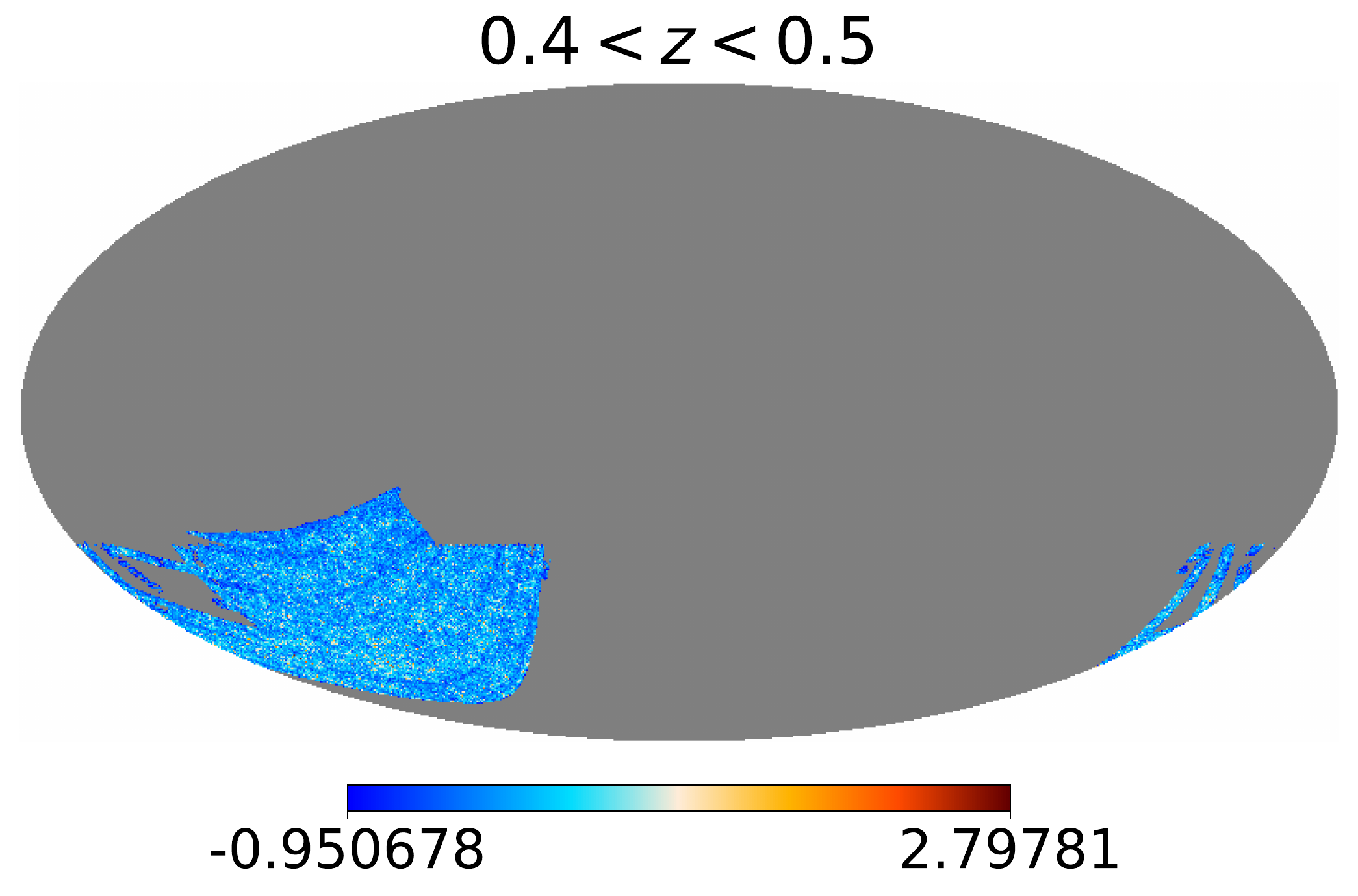}
    \end{subfigure}
    \begin{subfigure}[b]{0.475\textwidth}   
        \centering 
        \includegraphics[width=\textwidth]{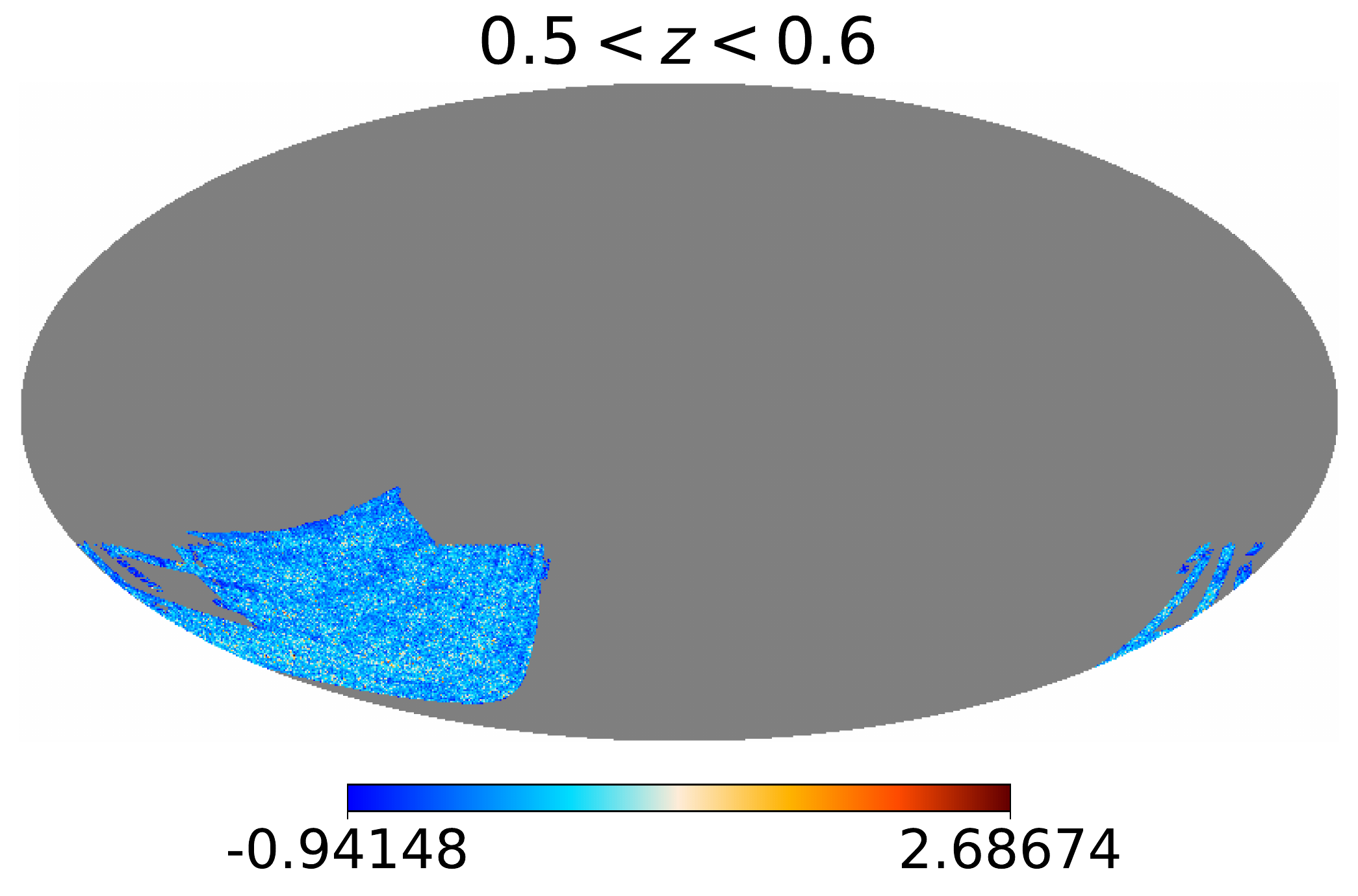}
    \end{subfigure}
    \begin{subfigure}[b]{0.475\textwidth}   
        \centering 
        \includegraphics[width=\textwidth]{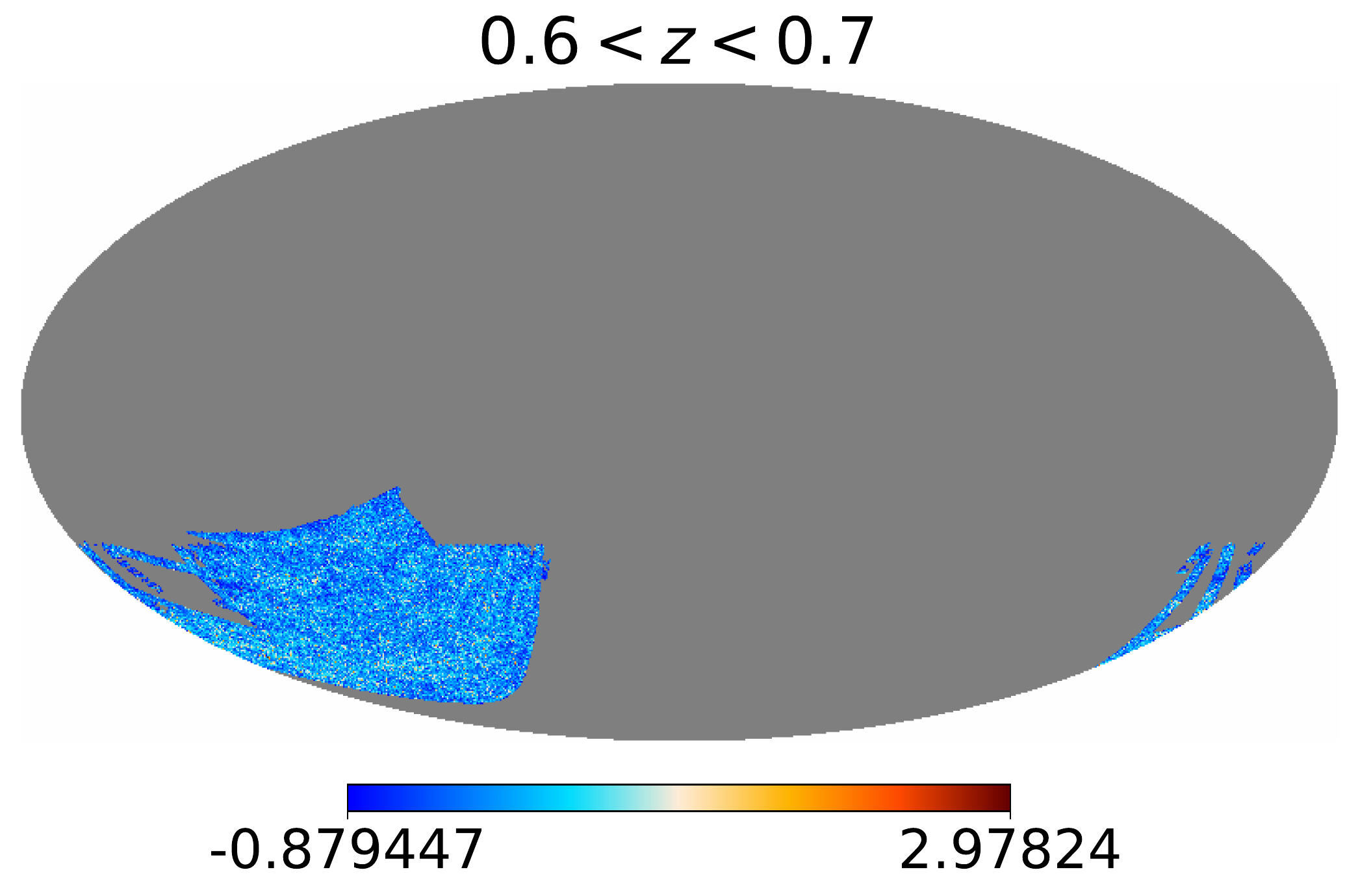}
    \end{subfigure}
    \caption{All-sky projections of the galaxy overdensity of the six photo-z bins adopted in the analysis. The maps are constructed in the \texttt{HEALPix} pixelization scheme, with the resolution parameter $N_{side} = 512$. The gray areas correspond to the masked regions.}
    \label{fig:tomography_delta}
\end{figure*}

In order to apply a tomographic approach, we split the full catalogue into six redshift bins of width $\Delta z =0.1$ over $0.1 < z < 0.7$. We ignore the extreme redshift bins where the fractional photo-z errors become large and the galaxy density became small. We use the position of the sources to create a pixelized overdensity map, 

\begin{equation}
\delta_{g}(\vec{x}) = \frac{n_{g}(\vec{x})- \bar{n}}{\bar{n}}, 
\end{equation}
where $n_{g}$ is the number of observed galaxies in a given pixel and $\bar{n}$ is the mean number of objects per pixel in the unmasked area. We use the \texttt{HEALPix} scheme \citep{gorski2005healpix} with a resolution parameter $N_{side}$= 512. The figure \ref{fig:tomography_delta} shows the overdensity map in these six redshift bins, where the 
gray area indicates the masked regions. However, we discard the stripes located in the galactic 
longitude range $180^{\circ} < l < 330^{\circ}$ due to the low density, remaining about $f_{sky}= 0.08$ 
in each map for analyses. 
The specifics of each bin are summarized in the table \ref{tab:density_specifications}.

\begin{figure}[htbp!]
	\centering
	\includegraphics[scale=0.5]{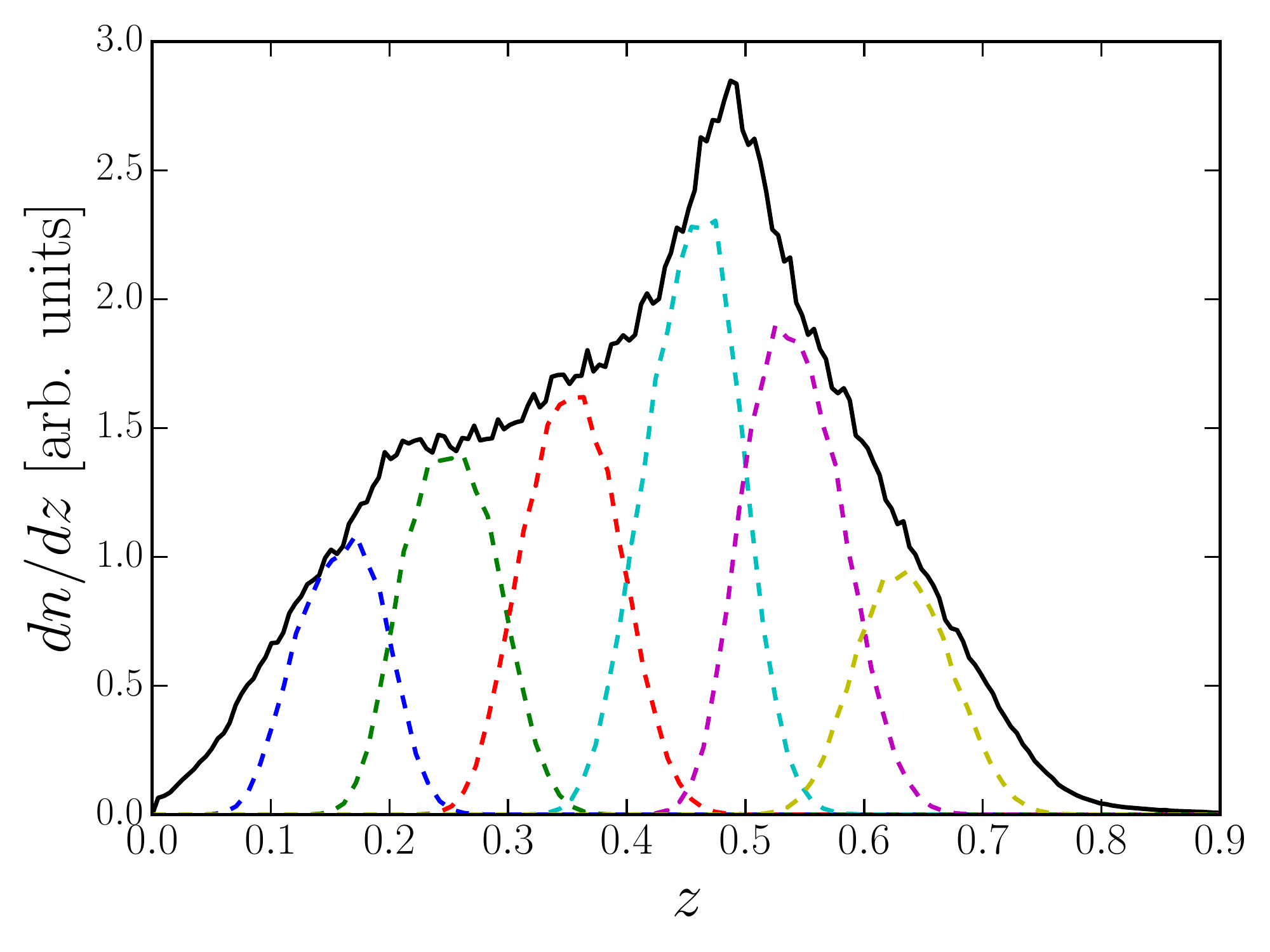}
	\caption{ Unnormalized redshift distributions for the total galaxy catalogue (black solid line) and for the six tomographic bins used in the analysis (dashed lines). These are obtained by convolving the photo-z distribution in each redshift interval with a photometric error distribution.}
\label{fig:dndz}
\end{figure}

As discussed in the section \ref{sec:theory}, we need the overall redshift distribution $dn/dz$ and the galaxy bias to connect the galaxy overdensity $\delta_{g}$ to the underlying matter overdensity $\delta$. However, we need take into account the effect of the photometric redshift errors \citep{budavari2003angular,sheth2010convolution}. We can accurately reconstruct the true $dn/dz$ distribution by the convolution of the sample's photometric redshift distribution $dn/dz(z_{ph})$ with the catalog's photo-z error function $p(z|z_{ph})$:
\begin{equation}
     \frac{dn}{dz} = \int_{0}^{\infty} dz_{ph} \frac{dn(z_{ph})}{dz} p(z|z_{ph})W(z_{ph}),
 \end{equation}
 where $ p(z|z_{ph})$ is parameterized as a Gaussian distribution with zero mean and dispersion $\sigma_z$ so that $p(z|z_{ph}) \propto \exp{(-0.5(z/\sigma_{z}(1+z))^{2})}$, where $\sigma_{z}= 0.019$~\citep{gao2018photometric} and the $W(z_{ph})$ is the window function, such that $W=1$ for $z_{ph}$ in the selected interval and $W=0$ otherwise. The redshift distribution for the total catalogue is shown as the solid black line in figure \ref{fig:dndz}, while the distribution to each tomographic bin is shown as the dashed lines. 
  
\begin{table}[ht]
\centering
\begin{tabular}{ccl}
\hline
Redshift range & $N_{tot}$ & $\bar{n}$ {[}gal $sr^{-1}${]} \\ \hline
0.1 - 0.2      & 2,208869  & $2.19 \times 10^{6}$              \\
0.2 - 0.3      & 3,178981  & $3.14 \times 10^{6}$              \\
0.3 - 0.4      & 3,686820  & $3.64 \times 10^{6}$              \\
0.4 - 0.5      & 5,155408  & $5.09 \times 10^{6}$              \\
0.5 - 0.6      & 4,348898  & $4.29 \times 10^{6}$              \\
0.6 - 0.7      & 2,101281  & $2.08\times 10^{6}$               \\ \hline
Total          & 20,680257 & $2.04\times 10^{7}$               \\ \hline
\end{tabular}
\caption{The number of sources and the galaxy number density of each tomographic bins.}
\label{tab:density_specifications}
\end{table}

\subsection{Planck CMB lensing}
We consider the latest CMB lensing products from \textit{Planck} 2018 data release \footnote{\href{http://pla.esac.esa.int/pla}{http://pla.esac.esa.int/pla}}. The lensing convergence map has been reconstructed based on the quadratic estimators that exploit the statistical information introduced by weak lensing in the CMB data \cite{okamoto2003cosmic}. Specifically, we use the convergence field $\kappa$ obtained from the minimum-variance (MV) combination of the estimators applied to temperature (T) and polarization (P) of the \texttt{SMICA} foreground-reduced map ~\citep{lensing2018planck}.

The spherical harmonic coefficients of the convergence is band-limited to the multipole $\ell_{max} =4096$. Jointly to the released lensing products, it is available a set of realistic simulations, which accurately incorporate the Planck noise levels and the $\kappa$ statistical properties \citep{aghanim2018plffp10}. In order to attenuate the foreground contamination in the lensing data, we apply the corresponding confidence mask that leaves a total unmasked sky fraction of $f_{sky}=0.671$.

\section{Method} 
\label{sec:method}
In this work, we use the angular power spectrum of the galaxy overdensity and the angular cross-power spectrum between the galaxy overdensity and the CMB convergence map to estimate the cosmic growth information at several redshifts bins. In this section, we describe the procedure followed in the analysis of these two datasets.

\subsection{Power Spectrum Estimation}
The power spectra estimates for incomplete sky coverage are affected by the mask, which introduces coupling between different modes \citep{hauser1973statistical}. Therefore, we use a pseudo-$C_{\ell}$ estimator based on the \texttt{MASTER} approach \cite{hivon2002master}, that provides a very good approximation to this issue, mainly on larger scales which is the regime we are considering, as detailed below.\ 

Let us denote the two fields $X$ and $Y$ with the auto-power spectrum when $X=Y$ and the full sky cross-(auto-)spectrum denoted as $C_{\ell}^{XY}$ ($C_{\ell}^{XX}$). The pseudo-$\tilde{C}_{\ell}^{XY}$ measured in a fraction of the sky is 
\begin{equation}
\tilde{C}_{\ell}^{XY} = \frac{1}{2 \ell +1} \sum_{m=-\ell}^{\ell} \tilde{X}_{\ell m} \tilde{Y}_{\ell m}^{*},
\end{equation}
where $\tilde{X}_{\ell m}$ and $\tilde{Y}_{\ell m}^{*}$ are the spherical harmonic coefficients of the maps. The mask acts as a weight modifying the underlying harmonic coefficients so that the pseudo-$C_{\ell}$ measured from the data can be related to the true spectrum by the mode-mode coupling matrix $M_{\ell \ell'}$ as
\begin{equation}
 \tilde{C}_{\ell}^{XY} = \sum_{\ell'}M_{\ell \ell'}C_{\ell}^{XY},
 \label{eq:coupling}
\end{equation}
where $M_{\ell \ell'}$ is inferred by the geometry of the mask \citep{hinshaw2003first}, given by

\begin{equation}
M_{\ell\ell'} = \frac{2\ell'+1}{4\pi}\sum_{\ell''}(2\ell''+1)\mathcal{W}_{\ell'} \begin{pmatrix}
  \ell & \ell' & \ell'' \\
  0 & 0 & 0 \\ 
 \end{pmatrix}^2 .
\end{equation}
Here $\mathcal{W}_{\ell'}$ is the angular power spectrum of the mask when $X=Y$, while in the cross-correlation corresponds to the two joint masks. In the cross-correlation analysis, we multiply the corresponding $\kappa$ and $\delta_{g}$ masks for each redshift bin.\

Depending on the size of the sky cut, the relation \ref{eq:coupling} cannot be inverted to obtain $C_{\ell}^{XY}$ because in general, the coupling matrix becomes singular. To mitigate the coupling effect and also to reduce the errors in the results, it is appropriate to bin the power spectrum in $\ell$. An unbiased estimator of the true-bandpowers $\hat{C}_{L}^{XY}$ is given in terms of the binned coupling matrix $K_{L L'}$
\begin{equation}
\hat{C}_{L}^{XY} =\sum_{L'\ell} K^{-1}_{LL'}P_{L'\ell}\tilde{C}_{\ell}^{XY},  
\end{equation}
where 
\begin{equation}
    K_{LL'}=\sum_{\ell\ell'}P_{L\ell}M_{\ell\ell'}B_{\ell'}^{X}B_{\ell'}^{Y}p^2_{\ell'}F_{\ell'}Q_{\ell'L'}.
    \label{eq:binned_coupling_m}
\end{equation}
Here $L$ denotes the bandpower index, $P_{L\ell}$ is the binning operator and $Q_{\ell'L'}$ is its reciprocal corresponding to a piece-wise interpolation. The $B_{\ell'}$ is a beam function for each $X$ and $Y$ observed field, $p_{\ell}$ is the pixel window function and $F_{\ell'}$ is the effective filtering function \citep{hivon2002master}.\
 
One of the advantages of the cross-correlation is that we do not need to debias the noise, since the CMB lensing and the galaxy data are completely independent measurements and therefore have, in principle, uncorrelated noise signals. However, we correct the galaxy power spectrum $\hat{C}^{gg}_{\ell}$ by subtracting the shot noise term: $N_{\ell}^{gg} =1/\bar{n}$, where $\bar{n}$ is the average number density of galaxies per steradian.

The analytical errors on the estimated auto-(cross-)spectrum are determined by \citep{balaguera2018extracting} 
\begin{equation}
     \Delta \hat{C}_{L}^{XY} = \sqrt{\frac{1}{(2L+1)f_{sky}\Delta\ell}}\big[(\hat{C}_{L}^{XY})^2+ \hat{C}_{L}^{XX}\hat{C}_{L}^{YY}\big]^{1/2},
\label{eq:errorcl}
 \end{equation}
where we assume in this equation that both fields behave as Gaussian random fields and the auto-power spectrum incorporates the associated noise, $N_{\ell}^{gg}$ and $N_{\ell}^{\kappa}$ for the galaxy and CMB lensing, respectively.\

We bin the power spectrum in $\ell$ in a linearly spaced band powers of width $\Delta\ell =10$ in the range $10 < \ell< 512$. We test different bin width values, however, we find no significant impact on the results. Due to the accuracy of the Limber approximation and the limited area covered by the survey, the power spectrum for $ \ell <20$ is poorly estimated, and we did not use it in our analysis. However, we include the first bin to perform the inversion of the binned coupling matrix of equation \ref{eq:binned_coupling_m} and the pseudo-$C_{\ell}$ calculation, to prevent the bias from the lowest multipole.

While we set the lowest value of $\ell$ on the $\ell_{min}= 20$, we impose a conservative cut in 
$\ell_{max}$ for each tomographic redshift bin to avoid several effects significant at small scales that could affect our results. From the theoretical galaxy power spectrum calculated by equation (\ref{eq:powerspe_th}), we have therefore limited our analysis to the scales where the percent deviation between the linear and non-linear models were smaller than $5\%$, corresponding to limit the analysis to modes $k_{max} \lesssim 0.07h$ Mpc$^{-1}$. We consider the same $\ell_{max}$ for the galaxy-galaxy and galaxy-CMB lensing analyses.

\subsection{$\hat{D}_{G}$ Estimator}

We need to take into account the errors associated with the power spectrum measurements to obtain the $\hat{D}_{G}$ properly \citep{giannantonio2016cmb,Bianchini2018}. Thus, we use the weighted average in the $\hat{D}_{G}$ calculation, 
\begin{equation}
    \hat{D}_{G} = \frac{\sum_{L}w_{L}\hat{D}_{G,L}}{\sum_{L}w_{L}}, 
    \label{eq:dg_estimator_avera}
\end{equation}
where the weights takes into account the variance in the $\hat{D}_{G}$ estimator
\begin{equation}
    w^{-1} = \hat{D}^2_{G,L}\bigg [\bigg(\frac{\Delta\hat{C}_{L}^{\kappa g}}{\hat{C}_{L}^{\kappa g}}\bigg)^2 +\frac{1}{4}\bigg(\frac{\Delta\hat{C}_{L}^{gg}}{\hat{C}_{L}^{gg}} \bigg)^2     \bigg],
\end{equation}
and the $\hat{D}_{G}$ per bandpower $L$ is written as
\begin{equation} 
    \hat{D}_{G,L}=\frac{\hat{C}_{L}^{\kappa g}}{\slashed{C}_{L}^{\kappa g}}\sqrt{\frac{\slashed{C}_{L}^{gg}}{\hat{C}_{L}^{gg}}}.
\end{equation}

\subsection{Covariance Matrix}
\label{subsec:covariance}
In order to check the possible impact of the off-diagonal contributions on the covariance matrix produced by non-linear clustering and by the mask as well as to verify the consistency of the error bars, we estimate the covariance with three approaches: analytical, jackknife (JK), and Monte-Carlo (MC) realizations.\

In the analytical approach, we assume Gaussianity of the fields and the covariance is simply diagonal with its elements computed by equation (\ref{eq:errorcl}). Although this method may be a good approximation when dealing with scales that are in the linear or mildly nonlinear regimes, it may be unrealistic to neglect the off-diagonal components of the covariance matrix \citep{lacasa2018covariance}. 

Hence, we calculate the JK technique by dividing the footprint covered by the masks into $N_{JK}$ regions, defined by the \texttt{HEALPix} pixelization scheme. We remove each region in turn and compute the power spectrum using the remaining subsample, such that the covariance is determined by
\begin{equation}
    Cov^{JK}(\hat{C}_{L}^{XY}, \hat{C}_{L'}^{XY}) = \frac{N_{JK}-1}{N_{JK}}\sum_{n=1}^{N_{JK}}(\hat{C}_{L, n}^{XY} -\bar{C}_{L}^{XY})(\hat{C}_{L', n}^{XY} -\bar{C}_{L'}^{XY}),
\end{equation}
where the $\hat{C}_{L, n}^{XY}$ is the power spectrum when removing the $n-$th jackknife region and the $\bar{C}_{L}^{XY}$ is the power spectrum averaged over all the jackknife regions. This method provides a covariance estimate in a model-independent way. We use $N_{JK}$ = 233 regions defined as \texttt{HEALPix} pixels with resolution $N_{side} = 16$. To establish this number we have considered as criteria the minimal number of $N_{JK}$ patches as those in which the scatter in the number of unmasked pixels deviates by less than $20\%$ from the mean, as described by \citep{balaguera2018extracting}. However, the results may depend on the number of masked-out regions and the region size. We have tested the results with different choices of the $N_{side}$ and we have found that starting at $N_{side} = 8$, the diagonal elements of the associated covariance matrix and the mean $\bar{C}_{L}^{XY}$ among the $N_{JK}$ regions are stable.\
 
\begin{figure}[!htbp]
    \centering
    \begin{subfigure}{1.0\textwidth}
        \centering
        \includegraphics[width=1.0\linewidth]{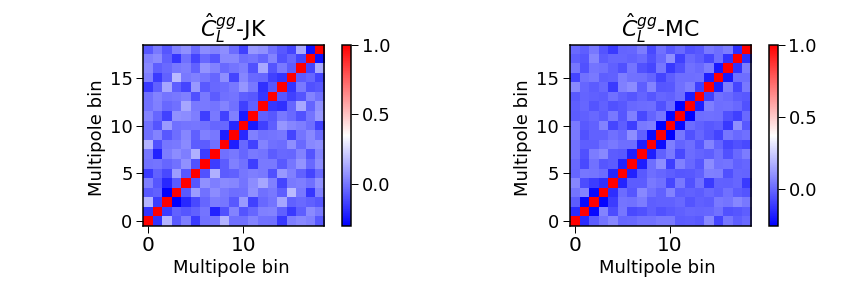}
    \end{subfigure}
    \centering
    \begin{subfigure}{1.0\textwidth}  
        \centering 
        \includegraphics[width=1.0\linewidth]{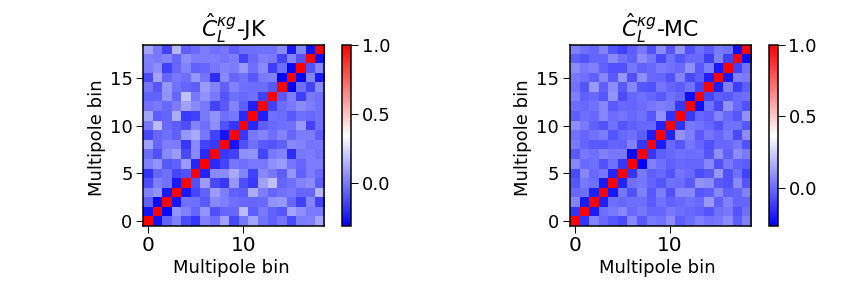}
    \end{subfigure}
    \caption{Covariance matrices normalized by their diagonal elements, obtained from different methods: Jackknife (left column) and Monte Carlo simulations (right column). In the first row, we show the covariance among the $C_{L}$ band-powers for the galaxy power spectrum and in the second row, we show the corresponding for the galaxy-CMB lensing cross-power spectrum. In both cases, we consider the full galaxy catalogue, spanning the redshift range $0.1<z<0.7$. }
    \label{fig:cov_comparison}
\end{figure}

Finally, we also have exploited MC simulations to build the covariance. For that, we generate 500 correlated Gaussian realizations~\citep{Copi:2016hhq,Kamionkowski:1996ks} of the galaxy and CMB lensing maps, considering their noise properties. To simulate the Gaussian convergence noise maps, we used the convergence noise power spectrum $N_{\ell}^{\kappa\kappa}$ provided by the Planck team~\citep{lensing2018planck}. From the galaxy mock, we generated a galaxy number count map assuming the galaxy number density of the data, where the value of each pixel is drawn as a Poisson distribution with the mean number of sources per pixel. We then, transform each number count map into a galaxy overdensity map and calculate the corresponding auto-power spectrum and cross-power spectrum using the lensing mock. The covariance evaluated from these measurements is
\begin{equation}
    Cov^{MC}(\tilde{C}_{L}^{XY}, \tilde{C}_{L'}^{XY}) = \frac{1}{N_{sims}-1}\sum_{j=1}^{N_{sims}}(\tilde{C}_{L, j}^{XY} -\langle\tilde{C}_{L}^{XY}\rangle)(\tilde{C}_{L', j}^{XY} -\langle\tilde{C}_{L'}^{XY}\rangle ).
    \end{equation}
In the figure \ref{fig:cov_comparison} we show the covariance matrix elements estimated for the JK method (left column) and for the MC method (right column), both normalized by their diagonal elements. In the first row are displayed the results for the galaxy $\hat C_{L}^{gg}$ and in the second row for the cross $\hat{C}_{L}^{\kappa g}$, both for the galaxy catalogue with the redshift range $0.1<z<0.7$. In the galaxy power spectrum, some off-diagonal elements are observed in both methods, although the amplitude of the off-diagonal elements is less than $14\%$ of the diagonal elements for the MC and $25\%$ for the JK matrix, considering the multipole range of $20< \ell< 250$. Regarding the cross-power spectrum covariance, the amplitude of the off-diagonal terms corresponds $\sim 10\%$ of the diagonal elements amplitude in the MC method and $\sim 15\%$ in the JK method. While for the MC these off-diagonal elements are assigned only due to the non-trivial correlations between angular multipoles added by the mask, in the JK matrices may incorporate also the non-Gaussian variance produced on small scales by the nonlinear evolution, the result is that the off-diagonal terms in the JK matrices are slightly larger than in the MC matrices.

\begin{figure}[!htbp]
        \centering
        \includegraphics[width=1.0\linewidth]{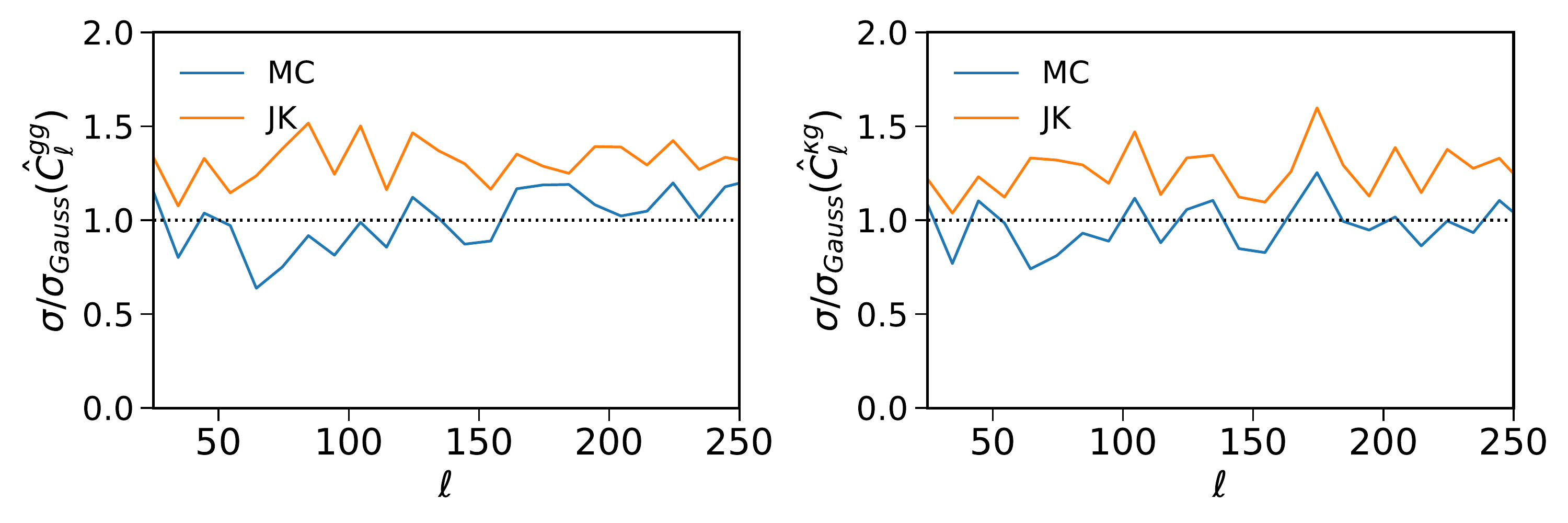}
        \caption{Comparison between the amplitude of the error bar estimated from the Jackknife covariance matrix and the analytical Gaussian (orange line) and from the Monte-Carlo covariance matrix and the analytical Gaussian (blue line). The results in the left column refer to the galaxy auto-power spectrum and, in the right, to the cross-power spectrum. }
    \label{fig:sig_comparison}
\end{figure}

To illustrate the comparison of the diagonal elements of the covariance matrices, we show in figure \ref{fig:sig_comparison} the errors from each matrices divided by the errors from the analytical estimate to the galaxy auto-power spectrum (left) and the cross-power spectrum (right). As expected, the amplitude of the MC errors is very similar to that of the Gaussian analytical errors as shown in the blue lines. In contrast, the JK errors are slightly larger than the analytical prediction, as shown in the orange lines. Therefore, in our analysis we adopt the JK method to be realistic in the error estimation and consistent in taking the off-diagonal terms into account, despite its low amplitude. We demonstrate in Appendix \ref{apd: covariance impact} that this choice does not have a significant impact on our main results for the six tomographic redshift bins of the galaxy catalogue.\ 
 
\section{Results}
 \label{sec:results}
We present here the measurements of the galaxy auto-correlation and the cross-correlation between the CMB convergence and galaxy overdensity. We also explore a number of checks carried out to ensure that our analysis is accurate and robust to observational and astrophysical systematic effects. Finally, we use our measurements to obtain the growth factor. 

\subsection{Galaxy bias and lensing amplitude}
\label{sec:galaxy_biasA}
\begin{figure}[!htbp]
    \centering
    \begin{subfigure}{.495\textwidth}
        \centering
        \includegraphics[width=0.95\linewidth]{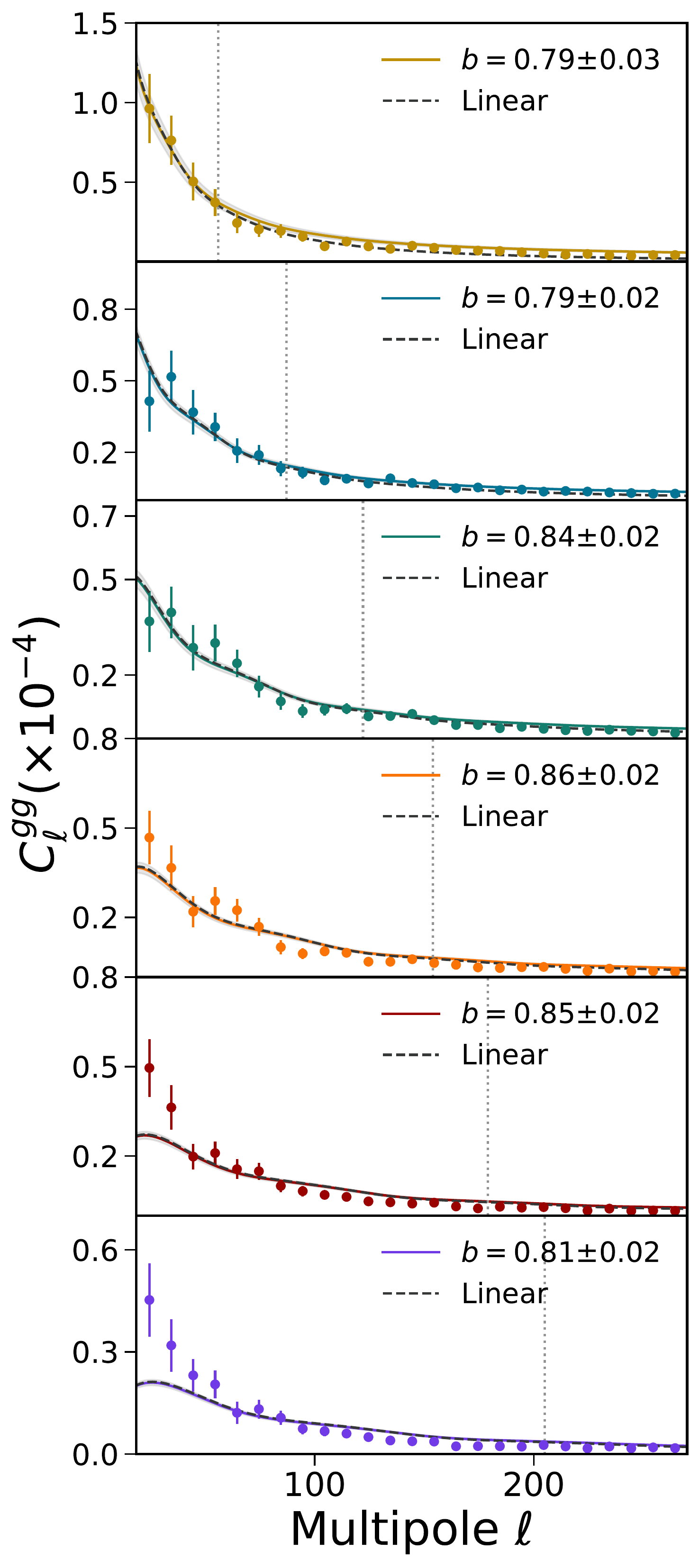}
    \end{subfigure}
    \centering
    \begin{subfigure}{.495\textwidth}  
        \centering 
        \includegraphics[width=0.95\linewidth]{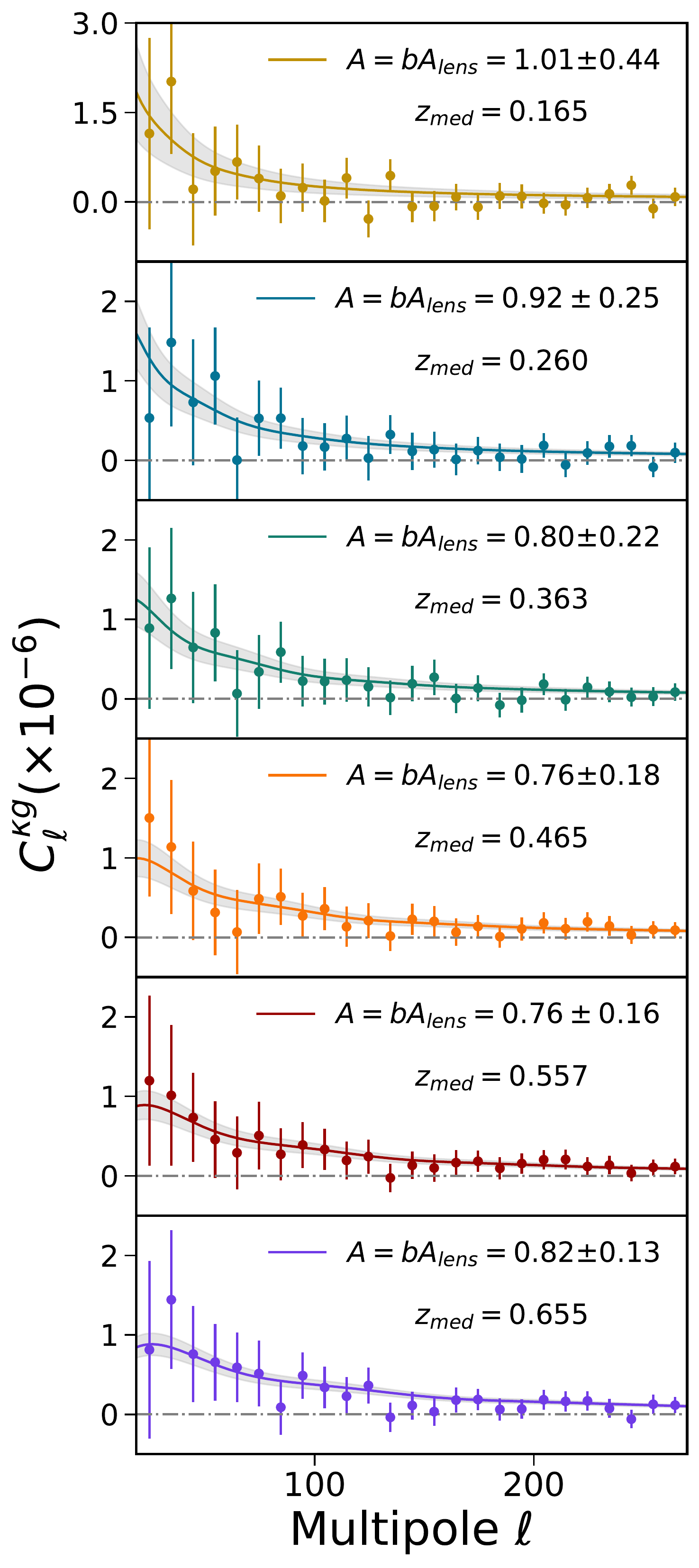}
    \end{subfigure}
    \caption{The galaxy auto-power spectrum (left panel) and the cross power spectrum (right panel) of the six tomographic redshift bins. The panels refer to the photo-z bins, from low to high redshift (top to bottom). {\red The median redshift of each bin is reported on each sub-panel of the cross power spectrum.} The points are the direct estimates while the solid (dashed) line is the fiducial cosmology including (excluding) the non-linear corrections, rescaled by the best-fit galaxy bias $b$ (for the auto-spectra) and by the cross-correlation amplitude $A= bA_{lens}$ (for the cross-spectra). The best-fit parameters are reported in the captions with their $1\sigma$ error. The shaded gray region represents the $1\sigma$ values around the best-fit model. The best-fit was inferred using up to the multipole $\ell_{max}$, represented as the vertical dashed line in the left panel}.  
    \label{fig:cls_data}
\end{figure}

We show the measurements of the $\ell$-binned galaxy power spectrum (left panel) and the cross-power spectrum (right panel) in Figure \ref{fig:cls_data}. The six panels represent, from top to bottom, the estimates to each redshift bin. The error bars are calculated using the JK covariance, as described in the previous section.\

Although the $\hat{D}_{G}$ estimator is bias-independent for a narrow redshift bin, we can use the observed galaxy-galaxy and the galaxy-CMB lensing power spectra to respectively estimate the best-fit bias $b$ and the amplitude of the cross-correlation $A= bA_{lens}$, where the later is introduced motivated by phenomenological reasons, and $A_{lens}$ is the CMB lensing amplitude ~\citep{ade2016isw}. Therefore, if the underlying cosmology conforms to the fiducial model, on average, it is expected $A_{lens}=1$ and consequently, the amplitude $A$ should assume the same value as the galaxy bias $b$ determined from the auto-correlation.

We assume that the bias does not evolve within each redshift bin so that A and b are free parameters obtained employing Bayesian analysis with an uninformative flat prior and a Gaussian likelihood,
 \begin{equation}
    \mathcal{L}(\mathbf{x}\mid \boldsymbol{\theta}) \propto \exp \left[-\frac{1}{2} \left(\mathbf{x}-\boldsymbol{\mu}\left(\boldsymbol{\theta}\right)\right)^T C^{-1} \left(\mathbf{x}-\boldsymbol{\mu}\left(\boldsymbol{\theta}\right)\right)\right],
\end{equation} 
where $\mathbf{x}$ is the extracted $\hat{C}_{\ell}^{gg}$ or $\hat{C}_{\ell}^{\kappa g}$, $\boldsymbol{\mu}$ is the correspondent binned theoretical prediction for the parameters $\boldsymbol{\theta}$, and $C^{-1}$ is the inverse of the covariance matrix of section \ref{subsec:covariance}. Following the conclusions of that section, we use the JK covariance matrix. When inverting the covariance matrix, we take into account the effect of having a finite number of realisations multiplying the inverse covariance by $(N_{JK}-p - 2) \,/\, (N_{JK} - 1)$ ~\citep{hartlap2007your}, where 
$p$ is the number of bins used.

In order to efficiently sample the parameter space, we use the Markov chain Monte Carlo (MCMC) method, employing \texttt{emcee}\footnote{\url{http://dfm.io/emcee/current/}} package \cite{foreman2013emcee}. 
We perform this analysis for each redshift bin and as a comparison, also for the full sample, spanning the redshift range $0.1 < z < 0.7$. Our results are stable against the length of the chain as well as the initial walker positions.\

The best-fit bias and the cross-correlation amplitude with their $1\sigma$ errors are reported in the captions of the Figure \ref{fig:cls_data}. The best-fit model with its $1\sigma$ uncertainties are shown as the solid lines and the gray shaded region, respectively. The vertical dashed line shown in the left panel indicates the $\ell_{max}$ multipole used in the analysis, where the nonlinear power spectrum differs from the linear theory by less than $5\%$. For the cross-power spectrum we use the same multipole range, however, we don't show it in the right panel for the sake of clarity. The significance of the parameter detection is calculated as $S/N =\sqrt{\chi^2_{null}-\chi_{min}^2(\theta)}$, where the $\chi^2_{null}$ is the $\chi^2(\theta=0)$ and $\chi_{min}^2(\theta)$ is the value for the best-fit. The parameter values, the $S/N$, the $\chi^2_{min}$, and the probability-to-exceed (PTE) for each redshift bin are summarized in Table \ref{table:bias}.\

 In the tomographic approach, we have found the best-fit bias in agreement up to $1\sigma$ with the cross-correlation amplitude, indicating the lensing amplitude consistent with unity. For all the redshift bins, the best-fit bias has $S/N$ greater than 11. Clearly, the cross-power spectrum constraints are less significant than those using the galaxy power spectrum, although we found $S/N \sim 1.53 - 5.61$. Finally, we see that in most cases the reduced $\chi^2$ is generally close to (or below) unity, indicating consistency of the fit, except for the 
{\red two highest redshift bins in the galaxy-galaxy estimate, where we found a poor PTE.\

The theoretical power spectrum defined by equation \ref{eq:powerspe_th} and consequently the quality of our constraints rely on the robustness of the distribution of the galaxies as a function of photometric redshift, $dn/dz$. Any incompatibility between the true and the assumed redshift distribution can potentially influence the inferred value of the parameters. Indeed, according to \cite{gao2018photometric}, for the galaxies with spectroscopic redshifts $z_{spec} \gtrsim 0.6$, the photometric redshifts of the catalogue tend to be underestimated mainly due to a lack of high-z galaxies in the training set and therefore, more susceptible to uncertainties in the modeling of the theoretical power spectrum in our analysis. As an example, we repeat the parameter constraints for the two highest redshift bins, $0.5<z<0.6$ and $0.6<z<0.7$, considering a broader and narrower $dn/dz$ distribution, taking $\sigma_{z} =0.04$ and $\sigma_{z} = 0.0$, respectively. We find that the galaxy-CMB lensing cross-correlations are extremely robust to the change of $\sigma_z$ due to the broadness of the CMB lensing kernel. However, the galaxy bias inferred from the galaxy-power spectrum is affected up to $\sim 30\%$, getting lower PTE and values when $\sigma_{z}=0.0$, being $b=0.71\pm 0.02$ (PTE= $0.71\%$) and $0.68\pm 0.02$ (PTE=$1.7\times10^{-4} \%$) for $0.5<z<0.6$ and $0.6<z<0.7$, respectively. In contrast, the best-fit galaxy bias assumes a higher value for $\sigma_{z} =0.04$, $b =1.08\pm0.03$ (PTE= 2.96$\%$) and $b =1.06\pm0.02$ (PTE = 0.11$\%$) for $0.5 < z < 0.6$ and $0.6 < z < 0.7$, respectively. It is important to emphasize that, an erroneous $\sigma_{z} $ was inserted only to exemplify how the fit of the galaxy bias may be affected if the photo-z's estimates are more biased at $z\gtrsim0.6$ than the considered in our $dn/dz$ distribution. Therefore, these results don't necessarily imply the presence of any systematic.\
}

\begin{figure}[!htbp]
    \centering
    \begin{subfigure}{0.49\textwidth}
        \centering
        \includegraphics[width=\linewidth]{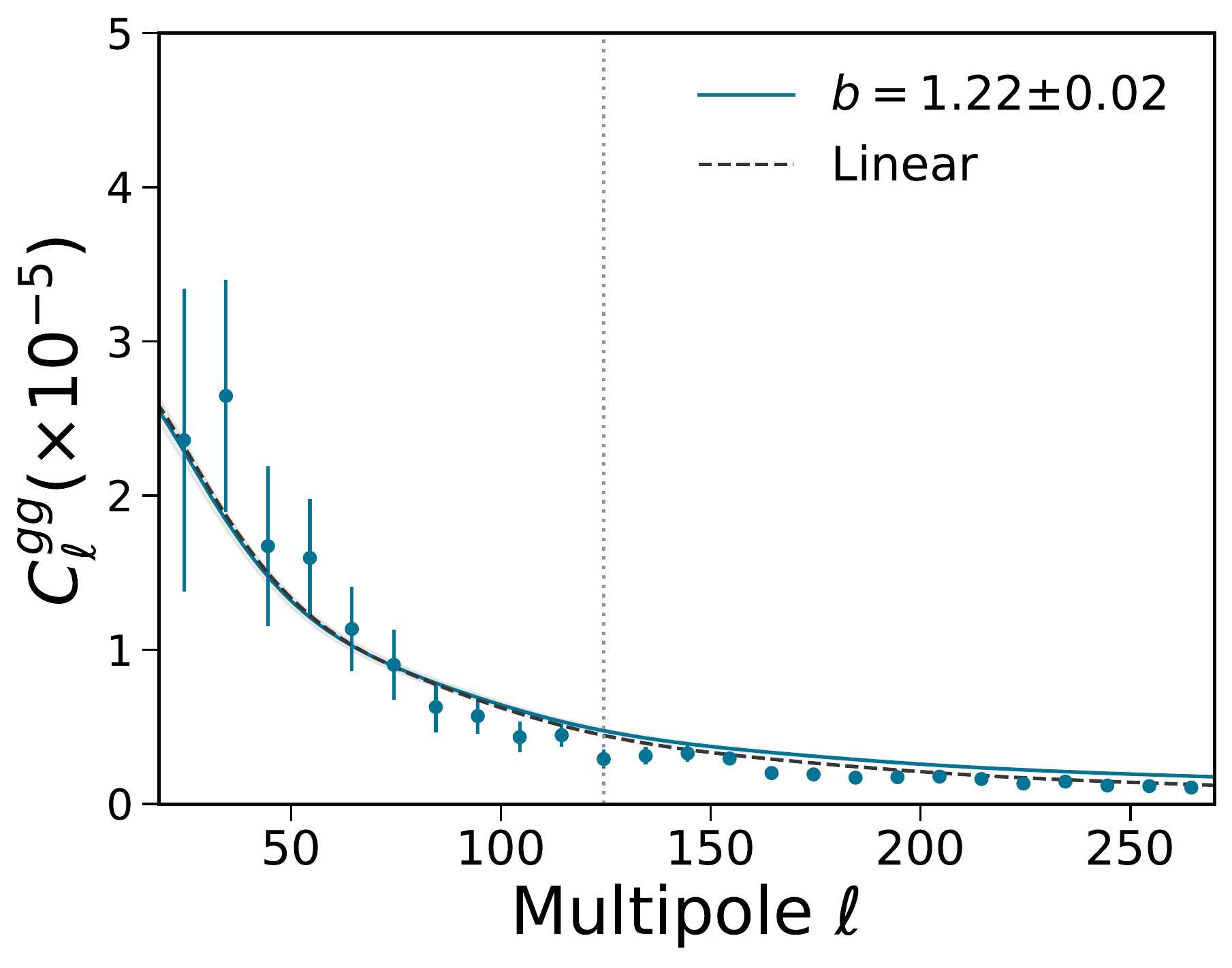}
    \end{subfigure}
    \centering
    \begin{subfigure}{0.49\textwidth}  
        \centering 
        \includegraphics[width=\linewidth]{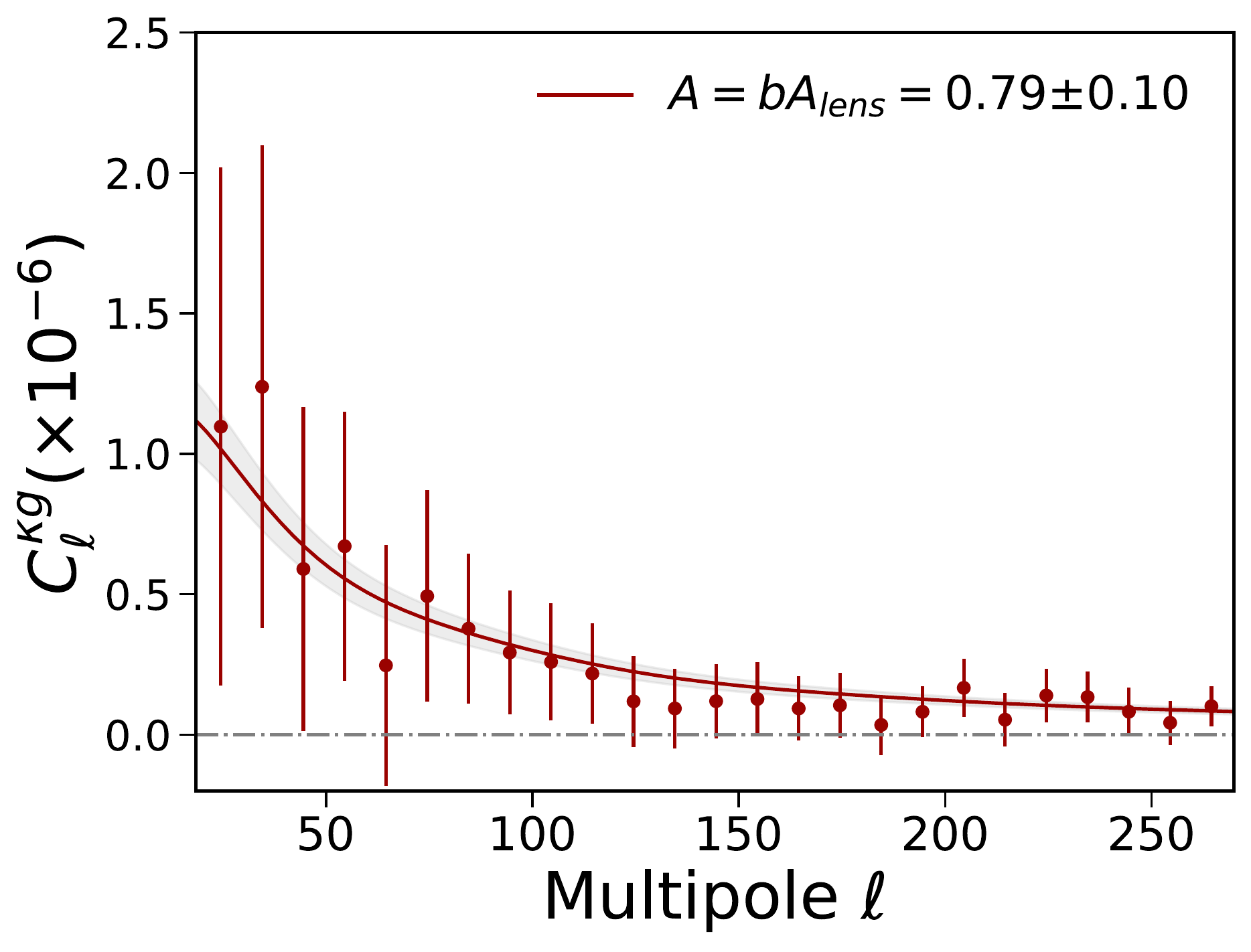}
    \end{subfigure}
    \caption{The galaxy auto-power spectrum (left panel) and the cross power spectrum (right panel) for the full galaxy sample in the redshift range $0.1 < z < 0.7$. As the Figure \ref{fig:cls_data}, the cross-correlation amplitude and the galaxy bias are reported in the captions with their $1\sigma$ error. The gray shaded region denote the $1\sigma$ values around the best-fit model (solid line)}. 
    \label{fig:cls_fullsample}
\end{figure}

The figure \ref{fig:cls_fullsample} shows the $\hat{C}_{\ell}^{\kappa g}$ (right panel) and the $\hat{C}_{\ell}^{gg}$ (left panel) when considering the sample of the galaxy covering the redshift from 0.1 to 0.7. We found that the galaxy auto-power is best fitted by our fiducial cosmology with linear galaxy bias $b=1.22\pm 0.02$. In contrast, we found that the cross-correlation with CMB lensing is best fitted by a lower amplitude value, $A<b$ in more than $3\sigma$ (including only statistical errors). The results are summarized also in table \ref{table:bias}. The reduced $\chi^2$ reveals that our estimate of the covariance is realistic and the model provides a good fit to the data. 

{\red To explore the possible reason for the trend $A <b $ found when using the whole catalog, but not found in the tomographic approach, we examine the variations of the result after remove galaxies located at high and low redshifts. When we discard the galaxies at $z= 0.7$ and $z=0.6$, in the bins $0.5 < z < 0.6$ and $0.6 < z < 0.7$, the difference between the bias $b$ and the cross-correlation amplitude $A$ is significantly reduced. Specifically, the inconsistency of $\sim 3.5\sigma$ found in $0.1<z<0.7$ decrease to $\sim 1.2\sigma$ (including only statistical errors) when we remove the galaxies in the bins $0.5<z<0.6$ and $0.6<z<0.7$, being $b= 1.00\pm 0.03$ and $A= 0.82\pm 0.11$. For the redshift range $0.1<z<0.4$, the A value is in agreement with $b$ within $1\sigma$, with $b= 0.91\pm0.02$ and $A=0.82\pm0.13$. However, when we discard only the galaxies at the lower redshifts, i.e., galaxies in the bins $0.1<z<0.2$ and $0.2<z<0.3$, the trend $A<b$ remains, with $b =1.20\pm 0.02$ and $A=0.78\pm0.10$ for the redshift range $0.2<z<0.7$ and $b =1.11\pm0.03$ and $A=0.76\pm0.11$ for $0.3<z<0.7$, respectively. Therefore, these results strongly suggest that the tension is driven by high redshift galaxies in the analyses. 
As a sanity check, we proceed with the analysis verifying the constraints of the growth function $\hat{D}_{G}(z)$ in section \ref{sec:constraints_dg} with and without including these high redshift z-bins.
}

{\red In addition to the photo-z uncertainties,} the inferred parameters may be changed by some effects, including the reddening and/or extinction and the foregrounds in the CMB maps. We consider in detail the impact of a variety of systematic effects and the null tests in Section \ref{sec:consistency}.\


\begin{table}[]
\begin{tabular}{cccccl}
\hline
Correlation      & Photo-z bin         & $b\pm \sigma_{b}$ & S/N   & $\chi^2/d.o.f$ & PTE ($\%$) \\ \hline
Gal-Gal          & $0.1 < z< 0.2$      & $0.79\pm 0.03$    & 11.21 & 0.21/3         & 97.51      \\
                 & $0.2 < z< 0.3$      & $0.79\pm 0.02$    & 17.16 & 6.02/6         & 42.09      \\
                 & $0.3 < z< 0.4$      & $0.84\pm 0.02$    & 20.48 & 11.56/9        & 23.92      \\
                 & $0.4 < z< 0.5$      & $0.86\pm 0.02$    & 25.59 & 25.31/12       & 1.34       \\
                 & $0.5 < z< 0.6$      & $0.85\pm 0.02$    & 30.45 & 32.27/15       & 0.59       \\
                 & $0.6 < z< 0.7$      & $0.81\pm 0.02$    & 34.83 & 50.58/19       & 0.006      \\
                 & full $0.1 < z< 0.7$ & $1.22\pm 0.02$    & 12.77 & 8.75/9         & 45.98      \\ \hline
Correlation      & Photo-z bin         & $A\pm \sigma_{A}$ & S/N   & $\chi^2/d.o.f$ & PTE($\%$)  \\ \hline
Gal- CMB lensing & $0.1 < z< 0.2$      & $1.01\pm 0.44$    & 1.53  & 1.01/3         & 79.79      \\
                 & $0.2 < z< 0.3$      & $0.92\pm 0.25$     & 2.87  & 3.24/6         & 77.77      \\
                 & $0.3 < z< 0.4$      & $0.80\pm 0.22$    & 4.19  & 3.18/9         & 95.64      \\
                 & $0.4 < z< 0.5$      & $0.76\pm 0.18$    & 3.12  & 2.88/13        & 99.63      \\
                 & $0.5 < z< 0.6$      & $0.76\pm 0.16$    & 5.42  & 1.82/15        & 99.99      \\
                 & $0.6 < z< 0.7$      & $0.81\pm 0.13$    & 5.61  & 4.96/19        & 99.89      \\
                 & full $0.1 < z< 0.7$ & $0.79\pm 0.10$    & 2.59  & 0.70/9         & 99.98      \\ \hline
\end{tabular}
 \caption{Summary of the results obtained from the galaxy-galaxy and galaxy-CMB lensing power spectra for the six redshift bins and for the full galaxy sample with redshift range $0.1 < z < 0.7$: the top half table shows the best-fit galaxy linear bias $b$ estimated from the galaxy auto-correlation, while the lower half shows the best-fit for the cross-correlation amplitude $A= bA_{lens}$. The signal-to-noise $(S/N)$ and the best-fit $\chi^2$ and the corresponding probability-to-exceed (PTE) are also shown.}
 \label{table:bias}
\end{table}

\subsection{Consistency and systematics tests}
\label{sec:consistency}
In this section, we summarise a number of tests carried out to ensure that our analysis is accurate and robust against systematic effects.

\subsubsection{Foreground contamination}
Although the CMB lensing map is reconstructed based on the \texttt{SMICA} foreground-reduced CMB map, there may still be residual contributions from galactic and extragalactic foregrounds and 
it can bias the lensing reconstruction and consequently the cross-correlation results. One of the largest potential contaminants is the thermal Sunyaev-Zel'dovich (tSZ) effect, which will also correlate with the galaxy density~\citep{van2014cmb,madhavacheril2018mitigating,geach2017cluster,schaan2019foreground}.\

\begin{figure*}[!htbp]
        \centering
        \includegraphics[width=0.6\textwidth]{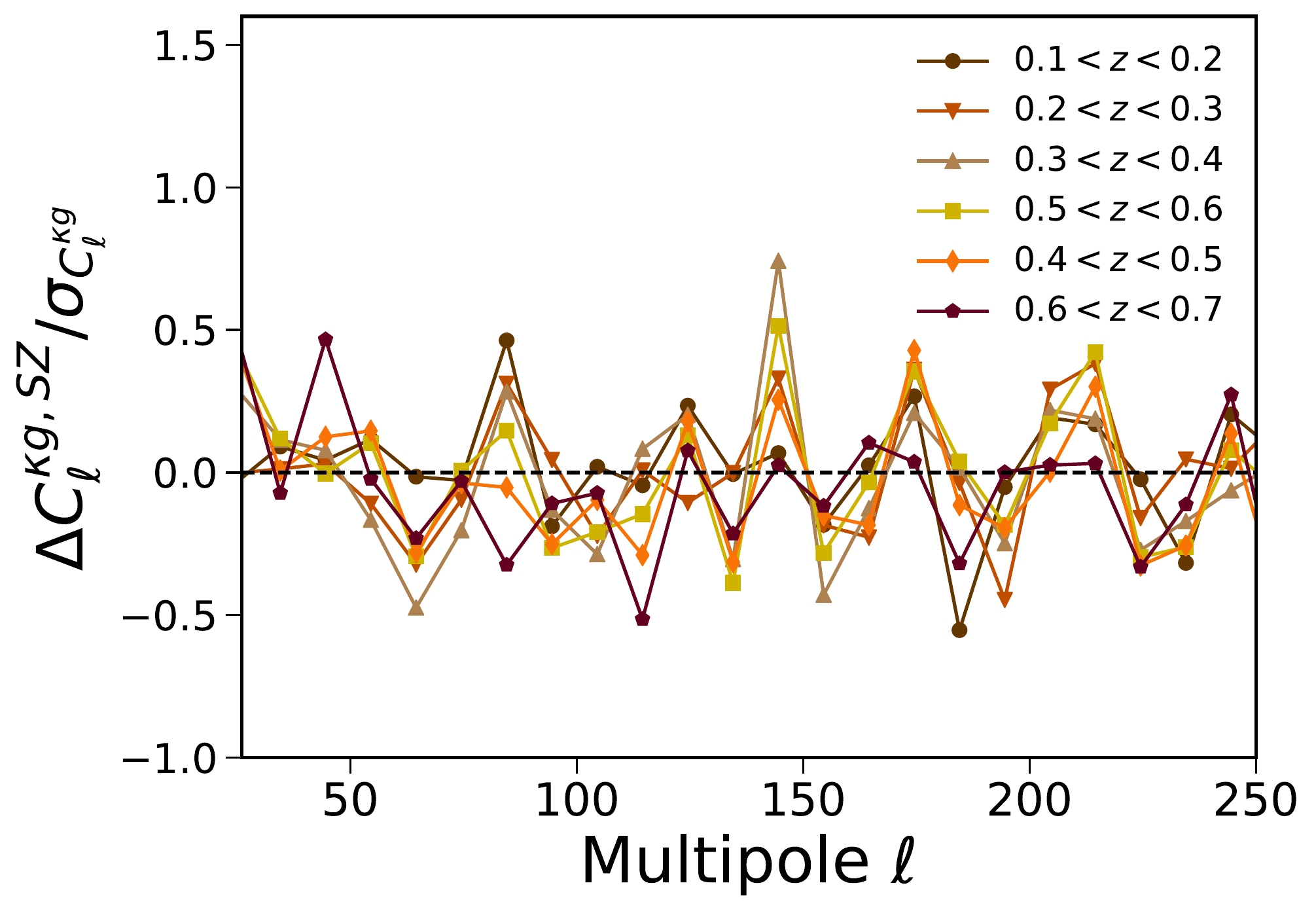}
        \caption{The difference of the galaxy-CMB lensing power spectrum when considering the CMB lensing map with the tSZ deprojected, in units of the statistical error. The several markers denote the result for each redshift bin. For all cases, the result does not deviate significantly compared to the statistical error amplitude.}
    \label{fig:sz_deproje}
\end{figure*}

We investigated the impact of the tSZ contamination by cross-correlating the galaxy overdensity map with the Planck CMB lensing reconstructed from the tSZ deprojected \texttt{SMICA} CMB map. Additionally, this test also allows us to check the Cosmic Infrared Background (CIB) bias since it is significantly different in the case of the tSZ-deprojected weighting~\citep{lensing2018planck}. In the figure \ref{fig:sz_deproje} we show the difference between the cross power spectrum estimation with and without the tSZ-deprojection, in units of the statistical error. For all redshift bins, we indeed see non-zero residuals coming from tSZ/CIB, and the removal of the tSZ contamination induces only sub-percentage changes in the cross-correlation, by less than $\sim 0.5\sigma$. Therefore, the consistency between the tSZ-free and the fiducial cross-correlations provides additional confidence that this foreground contamination does not affect significantly our overall results since it is subdominant with respect to other sources of uncertainty.\

\begin{figure*}[!htbp]
        \centering
        \includegraphics[width=0.6\textwidth]{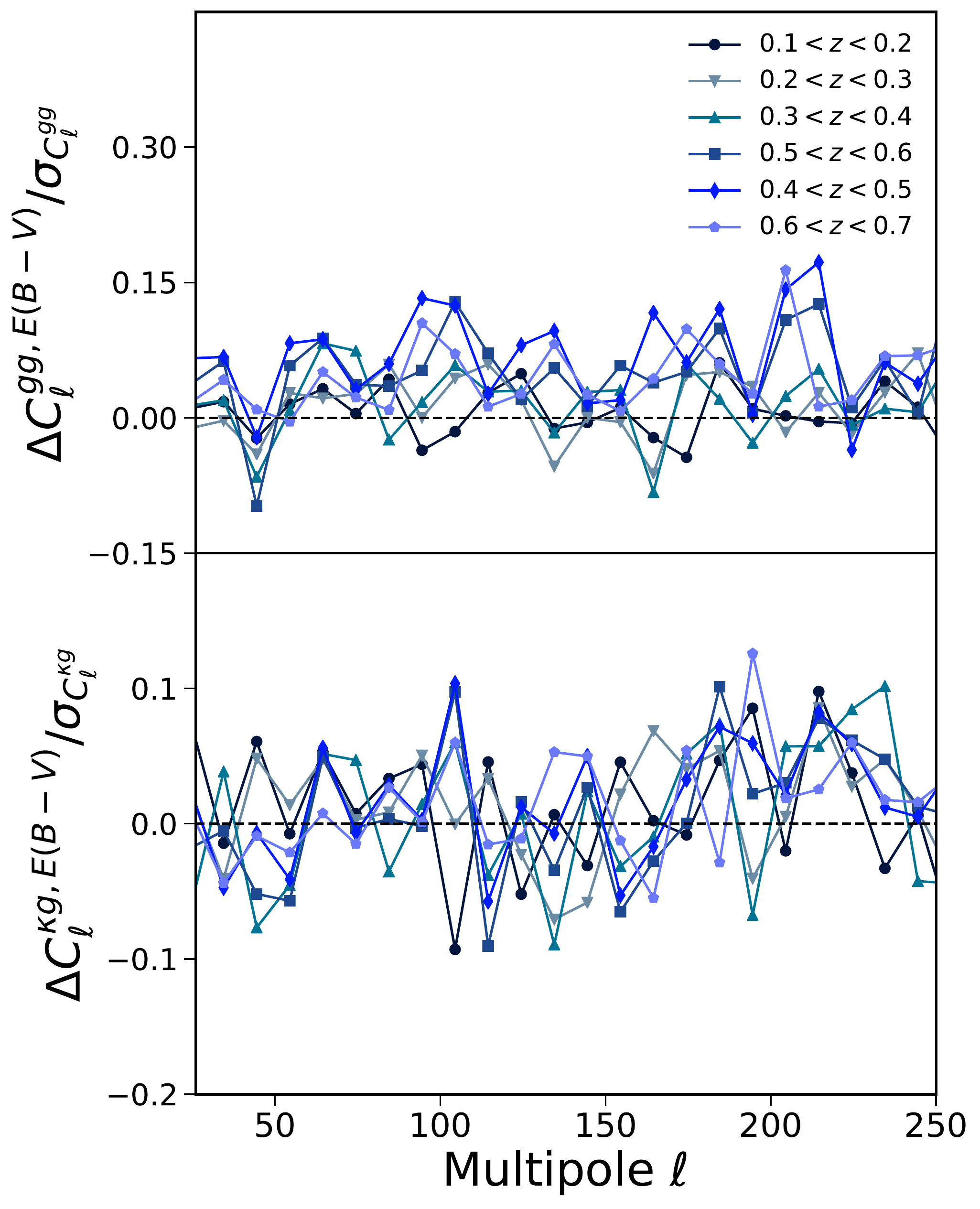}
        \caption{Stability of the measured galaxy power spectrum (top) and the galaxy-CMB lensing power spectrum (bottom), when applied the extinction correction, in units of the statistical error. We applied the extinction cut so that the most affected regions were removed, corresponding to $E(B-V)> 0.10$ of the Planck colour excess map.}
    \label{fig:reddening}
\end{figure*}

We also have tested the stability of the results with respect to the extinction of distant sources by dust in our galaxy. Although the galaxy magnitudes have already been corrected by the reddening map of \cite{schlegel}, we additionally use the extinction data from the Planck colour excess map \citep{abergel2014planck}. We create a mask in a \texttt{HEALPix} scheme, excluding all pixels for which the extinction map has $E(B-V) > 0.10$ mag, removing about $15\%$ of the pixels. We set this limit by looking at the contribution of $E(B-V)$ along the pixels within the galaxy footprint, verifying that the most dusty areas are in this range. We then, use this mask jointly with the galaxy and CMB lensing masks to calculate the power spectrum. In the figure \ref{fig:reddening} we show for the galaxy power spectrum (upper panel) and the cross power spectrum (lower panel) the difference of the results when we consider the extinction mask, in units of the statistical error. For both cases, we can see that the results are stable against the extinction correction, with variations within the statistical errors by less than $0.15\sigma$.\

\subsubsection{Null tests}

\begin{figure*}[!htbp]
    \centering
    \begin{subfigure}[b]{0.49\textwidth}
        \centering
        \includegraphics[width=\textwidth]{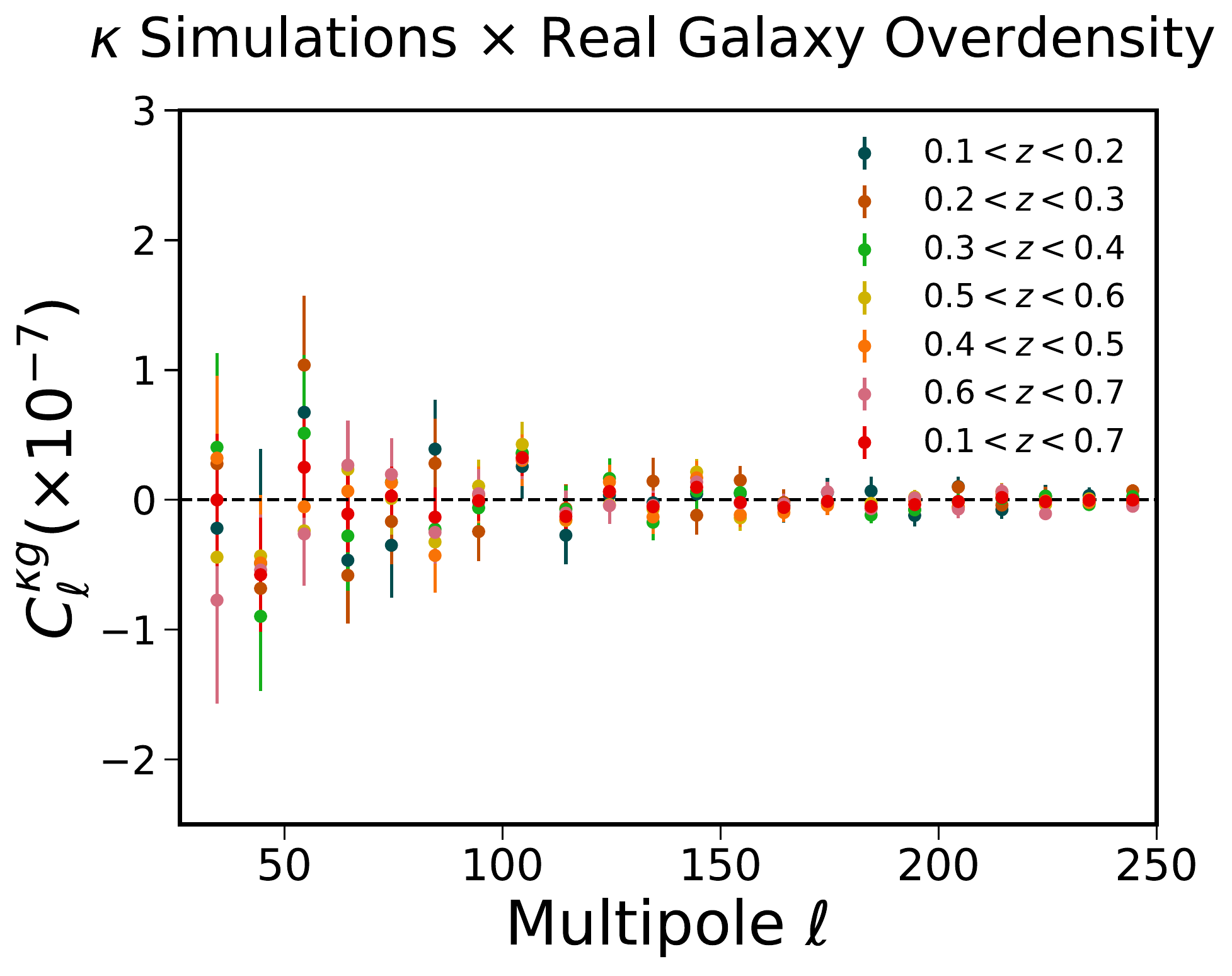}
    \end{subfigure}
    \centering
    \begin{subfigure}[b]{0.49\textwidth}  
        \centering 
        \includegraphics[width=\textwidth]{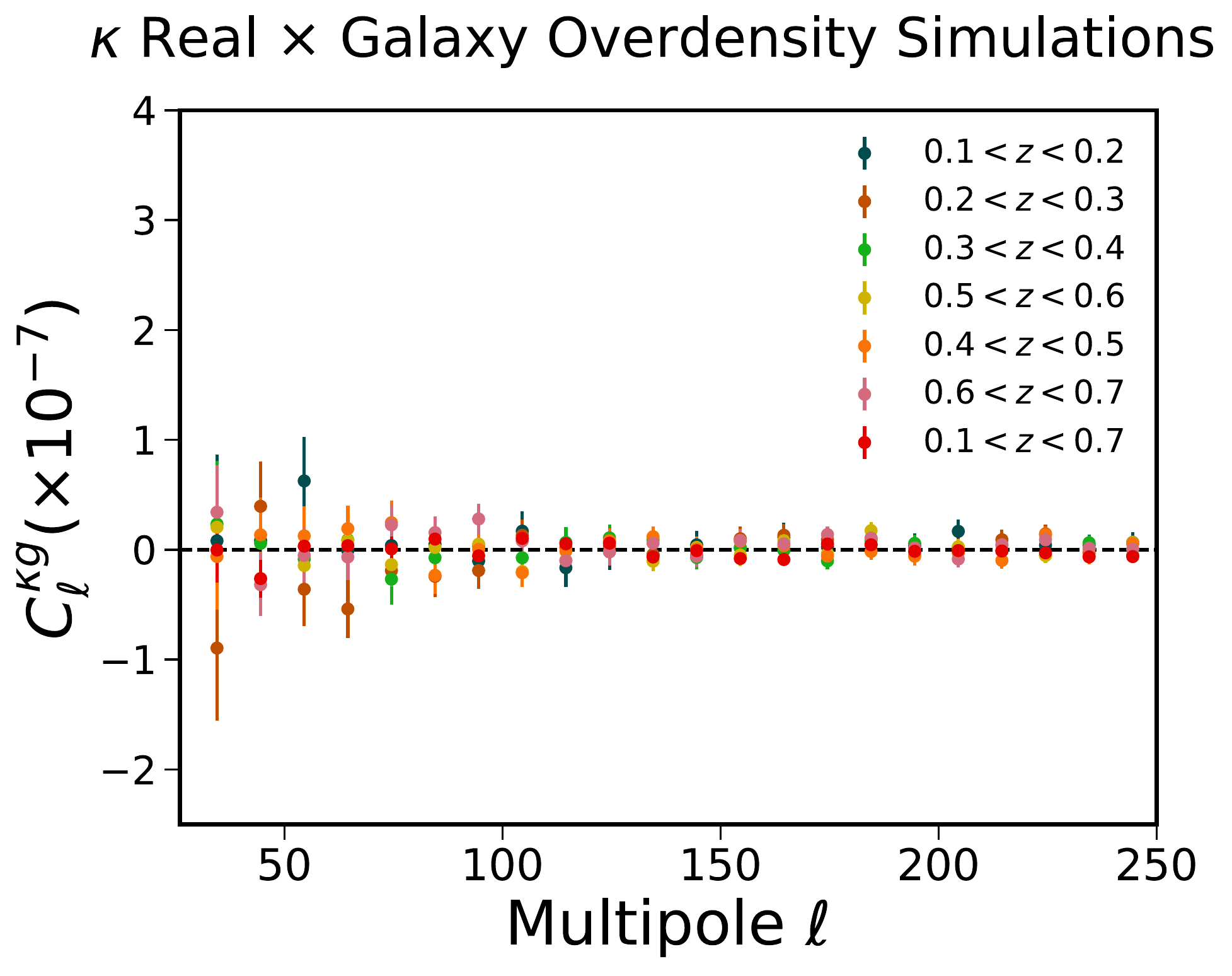}
    \end{subfigure}
    \caption{Null tests for the cross-power spectrum to the six redshift bins. In the right panel, is the mean correlation between the Planck CMB convergence map and 300 galaxy overdensity simulations obtained considering the respective features of each redshift bin. In the left panel, is the mean correlation between the galaxy overdensity and the 300 simulated Planck CMB lensing maps. The errors bars are given by the square root of the covariance matrix diagonal derived from the set of simulations and divided by $\sqrt{300}$.}
    \label{fig:null}
\end{figure*}

\begin{table}[!htbp]
\centering
\begin{tabular}{cccc}
\hline
Correlation                       & Photo-z bin    & $\chi^2$/$d.o.f$ & PTE ($\%$) \\ \hline
$\kappa$ Planck Sims $\times$ Gal & $0.1 < z< 0.2$ & 1.86/4           & 76.14      \\
                                  & $0.2 < z< 0.3$ & 10.29/7          & 17.26      \\
                                  & $0.3 < z< 0.4$ & 8.97/10          & 53.43      \\
                                  & $0.4 < z< 0.5$ & 12.82/13         & 46.17      \\
                                  & $0.5 < z< 0.6$ & 18.83/16         & 27.74      \\
                                  & $0.6 < z< 0.7$ & 14.48/20         & 80.51      \\
\multicolumn{1}{l}{}              & $0.1<z<0.7$    & 9.24/10          & 50.94      \\ \hline
Correlation                       & Photo-z bin    & $\chi^2/d.o.f$         & PTE ($\%$) \\ \hline
$\kappa$ Planck $\times$ Gal Sims & $0.1 < z< 0.2$ & 2.52/4           & 64.00      \\
                                  & $0.2 < z< 0.3$ & 10.17/7          & 17.89      \\
                                  & $0.3 < z< 0.4$ & 2.09/10          & 99.55      \\
                                  & $0.4 < z< 0.5$ & 10.01/13         & 69.35      \\
                                  & $0.5 < z< 0.6$ & 11.36/16         & 78.61      \\
                                  & $0.6 < z< 0.7$ & 22.92/20         & 29.25      \\
\multicolumn{1}{l}{}              & $0.1<z<0.7$    & 12.80/10         & 23.47      \\ \hline
\end{tabular}
\caption{ Summary of $\chi^2$ and the PTE for the null tests. The top half of the table shows the results for the cross-correlation between the Planck CMB lensing simulations and the real galaxy map, while the lower half shows the corresponding values for the real CMB lensing map correlated with the galaxy density simulations.}
\label{tab:null}
\end{table}

In order to check the validity of the cross-correlation against the possibility of residual systematics or spurious signals in the data, we perform a null hypothesis test of no correlation between the CMB lensing and the galaxy density maps. We do this by considering the cross-correlation of these two fields, being one of them the real map and the second one from simulations. As these maps do not contain a common cosmological signal, the mean correlation is expected to be consistent with zero.\

For each redshift bin, we cross-correlate the real galaxy maps with the $300$ convergence simulations from the Planck 2018 data release \citep{lensing2018planck}. In addition, we cross-correlate the Planck CMB convergence map with $300$ galaxy simulations constructed considering the same properties of the real galaxy data such as the best-fit bias, masks, and galaxy number density. The Figure \ref{fig:null} shows the cross-power spectrum estimated in both cases, where the errors bars were computed by the standard deviation of the simulated cross-power spectra divided by the $\sqrt{N_{sim}}$, with $N_{sim}= 300$.\

Considering the covariance matrices obtained from these simulations, we calculate the $\chi^2$ 
and the PTE. The results are displayed in the Table~\ref{tab:null}. We conclude that no significant signal is detected in either case and therefore, our cross power spectrum measurements are robust. 

\subsection{Constraints of $\hat{D}_{G}$}
\label{sec:constraints_dg}
\begin{figure}[!ht]
 \centering
     \includegraphics[scale=0.5]{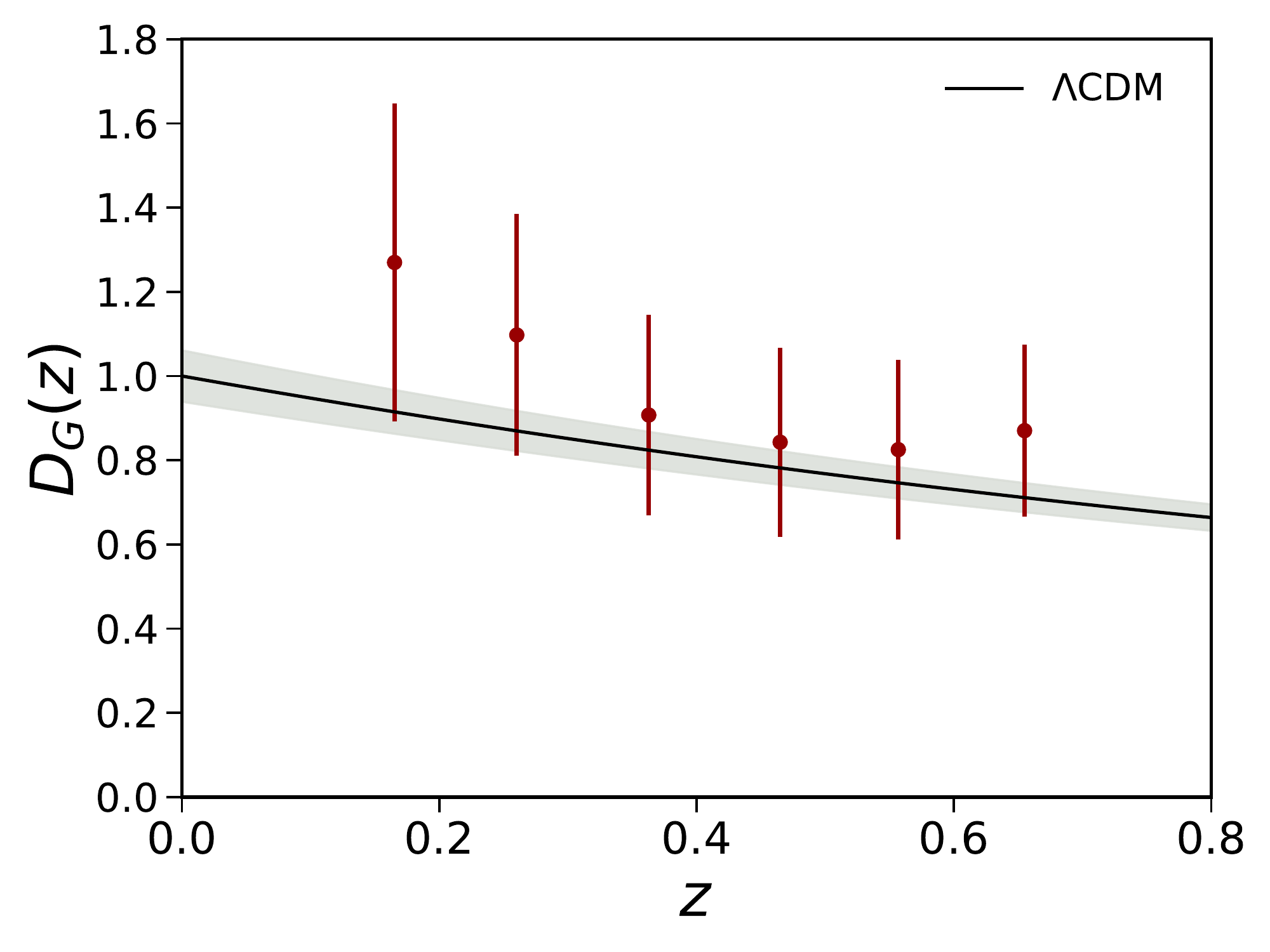}
     \caption{The linear growth factor estimated from the $\hat{D}_{G}$ estimator to the six redshift bins. The solid black line denotes the theoretical growth function for the Planck fiducial cosmology. The $2\sigma$ scatter for $3000$ cosmologies randomly drawn from the Planck chains is shown in the gray shaded region.}
    \label{fig:dgs}
 \end{figure}

We apply the $\hat{D}_{G}$ estimator, defined in the equation \ref{eq:dg_estimator_avera}, to the extracted bandpowers of the datasets. The figure \ref{fig:dgs} shows the growth factor for each redshift bin with the corresponding $1\sigma$ error bar. The error bars are estimated from the dispersion of the $\hat{D}_{G}^{sim}$, established from auto- and cross- spectra of the 500 correlated MC Gaussian realizations performed in the section \ref{subsec:covariance}. The solid black line denote the curve expected in the fiducial Planck $\Lambda$CDM model, $D_{G}^{fid}(z)$. As the function $D_{G}^{fid}(z)$ is directly related to the cosmological parameters $\Omega_{m}\sigma_{8}H_{0}^2$, we consider the Planck chains to randomly draw 3000 points and calculate the linear growth function for each cosmology. The gray shaded region around the $D_{G}^{fid}(z)$ indicates the $2\sigma$ scatter for the 3000 cosmologies. It is worth mention that for each cosmology $i$, we normalize the curve by multiplying by the factor ($\Omega_{m}\sigma_{8}H_{0}^2)^{i}$/($\Omega_{m}\sigma_{8}H_{0}^2)^{fid}$.\

We can assess the amplitude of the linear growth function $A_{D}$, with respect to the fiducial prediction, assuming a template shape of the $D_{G}(z)$ to be fixed by the $D_{G}^{fid}(z)$ \citep{giannantonio2016cmb,Bianchini2018,omori2018dark}, such that 
\begin{equation}
    D_{G}(z) =A_{D}D_{G}^{fid}(z).
    \label{eq:fit_parametrized}
 \end{equation}
The result of the fit is $A_{D} = 1.16 \pm 0.13$, when we consider the $\hat{D}_{G}$ of the six  redshift bins. 
{\red Due to the features of the highest redshift bins reported in section \ref{sec:galaxy_biasA}, as a sanity check we repeat the $A_{D}$ fit using only the redshift bins that have a robust bias fit, that is, only the three lowest redshift bins}, from which we obtain $A_{D} = 1.22\pm 0.19$. In both cases, we found consistency with the fiducial value, only slightly higher than the expected value $A_{D}^{\Lambda \mbox{\footnotesize CDM}} =1$.
  
Similar analyses using other galaxy samples, for nearer \citep{Bianchini2018,peacock2018wide} and for deeper \citep{giannantonio2016cmb,omori2018dark} redshifts than the considered in this work, also indicate agreement with the fiducial cosmology established by Planck. In this sense, our analysis is complimentary, as we consider another survey that covers a different region of the sky and therefore, extends to probing other possible systematics effects and redshifts intervals.
 
\section{Conclusions}
\label{sec:conclusion}
Recent reports witness the increasing importance of measuring the growth of the cosmic structures using the large deep surveys catalogues and CMB lensing data \cite{giannantonio2016cmb,peacock2018wide, Bianchini2018,omori2018dark}, now available. In fact, the linear structure growth factor as a function of redshift, $D_{G}(z)$, have the potential to discriminate between alternative models of cosmic acceleration. In this work, we present a tomographic estimate of the linear growth factor by combining the auto- and cross-correlation of the CMB Planck convergence map, $\kappa$, and a galaxy density fluctuations map, $\delta_g$, where the $\delta_{g}$ map was constructed from the photometric catalogue based on multi-band data from SCUSS, SDSS, and WISE \citep{gao2018photometric}. We perform detailed analyses in six redshift bins of width $\Delta z =0.1$, spanning the redshift interval $0.1 < z < 0.7$. 

We have studied the evolution of the linear galaxy bias, $b$, and the amplitude of the cross-correlation, $A$, using the galaxy-galaxy and galaxy-CMB lensing power spectra, respectively. We found a significant detection of the best-fit parameters, {\red with SN spanning $\sim 11-30\sigma$ for $b$ and $\sim 1.5- 5.6\sigma$ for $A$. However, we have found a poor fit for $b$ in the highest two redshift bins, possibly due to a lack of high-z galaxies in the training set during the construction of the photometric redshift catalogue \cite{gao2018photometric}, which consequently generate an underestimated photometric redshifts for $z\gtrsim 0.6$, increasing the uncertainties of $dn/dz$ in these redshifts bins. We have found the linear galaxy bias is in agreement with the cross-correlation amplitude and, therefore, with the lensing amplitude, in all redshift bins. When we consider the catalogue in the redshift range $0.1<z<0.7$ instead of a tomographic approach to fit the parameters, we found a trend of $A <b$, although the difference between them is reduced significantly when using galaxies only up to $z<0.6$, suggesting that this discrepancy is driven by the high redshift. The main results are summarized in Table \ref{table:bias} and section \ref{sec:galaxy_biasA}. }
 
In addition, we perform null tests to check if our measured signal is  affected by artifacts from the survey's systematics or other undesirable effects. To this end, we performed various investigations of the robustness of the results, showing that the null test indicates that the cross-correlation is unlikely to be affected by such effects and a possible tSZ and extinction contamination are negligible effects. These main results are displayed in Figure \ref{fig:sz_deproje}, \ref{fig:reddening} and \ref{fig:null}.\

By combining the auto and the cross-correlation estimates, we measure the linear growth factor at different epochs of the Universe by using the bias-independent estimator $\hat{D}_{G}$ introduced by \cite{giannantonio2016cmb}. The main result displayed in Figure~\ref{fig:dgs}, shows the measured linear structure growth factor in comparison with the expected in the fiducial $\Lambda$CDM scenario. Compiling the measurements of growth in the six tomographic bins, we find the amplitude of the linear growth function $A_{D}=1.16 \pm 0.13$, closely consistent with the expected by the fiducial model $A_{D}^{\Lambda \mbox{\footnotesize CDM}} = 1$.

The CMB lensing tomography is an efficient method to test the linear growth of cosmic structures and, by extension, to test dark energy scenarios and/or alternative gravity models. In the near future, the CMB and galaxy surveys such as the Simons Observatory, CMB-S4, LSST, and WFIRST will produce comprehensive data and will enable us to reach a deep mapping of the galaxies and a high sensitivity in reconstructing the CMB lensing potential. Thus, we may expect that the CMB lensing tomography, through analysis as the one used here, will be fundamental to find shrunken bounds in the scenario that better explains the history of the cosmic structure growth. 
\acknowledgments 
We thank Carlos Bengaly and Kevin Huffenberger for many useful discussions. GAM is supported by the CAPES Foundation of the Ministry of Education of Brazil fellowships. AB acknowledges a CNPq fellowship. We acknowledge the use of public data ~\cite{ade2016xv,gao2018photometric} and the use of many python packages: Numpy \cite{oliphant2015guide}, Astropy\footnote{http://www.astropy.org} a community-developed core Python package for Astronomy~\cite{astropy:2013, astropy:2018}, Matplotlib~\cite{hunter2007matplotlib}, IPython~\cite{perez2007ipython} and Scipy~\cite{jones2001scipy}.

\bibliographystyle{JHEP}
\bibliography{dg_paper}

\appendix

\section{Covariance matrix validation}
\label{apd: covariance impact}
As described in sec.\ref{subsec:covariance}, we use in our analysis the covariance matrix based on the jackknife method (JK). Here we present a additional test for the covariance matrix choice by calculating the bias and the cross-correlation amplitude of the six redshift bins, considering the Monte-Carlo (MC) covariance rather than the JK covariance.\

We summarise in table \ref{tab:mc_cons} the best-fit results obtained using the MC covariance. We can see that the agreement between the methods is excellent for all redshift bins, since the best-fit results are consistent with the findings of table \ref{table:bias}. Thus, we conclude that the choice of covariance does not significantly affect our results.

\begin{table}[h!]
\begin{tabular}{cccccl}
\hline
Correlation      & Photo-z bin    & $b\pm \sigma_{b}$ & S/N   & $\chi^2/d.o.f$ & PTE ($\%$) \\ \hline
Gal-Gal          & $0.1 < z< 0.2$ & $0.79\pm 0.04$    & 11.21 & 0.21/3         & 97.38      \\
                 & $0.2 < z< 0.3$ & $0.79\pm 0.03$    & 17.15 & 6.35/6         & 38.51      \\
                 & $0.3 < z< 0.4$ & $0.85\pm 0.03$    & 20.49 & 10.89/9        & 28.33      \\
                 & $0.4 < z< 0.5$ & $0.86\pm 0.02$    & 25.60 & 24.56/12       & 1.70       \\
                 & $0.5 < z< 0.6$ & $0.85\pm 0.02$    & 30.47 & 30.99/15       & 0.88       \\
                 & $0.6 < z< 0.7$ & $0.83\pm 0.02$    & 36.49 & 51.03/19       & 0.009      \\ \hline
Correlation      & Photo-z bin    & $A\pm \sigma_{A}$ & S/N   & $\chi^2/d.o.f$ & PTE($\%$)  \\ \hline
Gal- CMB lensing & $0.1 < z< 0.2$ & $0.99\pm 0.41$    & 1.54  & 1.00/3         & 80.10      \\
                 & $0.2 < z< 0.3$ & $0.99\pm 0.26$    & 2.85  & 3.37/6         & 75.99      \\
                 & $0.3 < z< 0.4$ & $0.80\pm 0.22$    & 4.18  & 3.25/9         & 95.35      \\
                 & $0.4 < z< 0.5$ & $0.72\pm 0.17$    & 4.87  & 3.54/13        & 99.03      \\
                 & $0.5 < z< 0.6$ & $0.78\pm 0.16$    & 5.41  & 1.91/15        & 99.99      \\
                 & $0.6 < z< 0.7$ & $0.83\pm 0.27$    & 6.00  & 3.85/19        & 99.99 \\   \hline 
\end{tabular}

\caption{Summary of the results for the galaxy-galaxy and galaxy-CMB lensing power spectrum for the MC covariance matrix. The results are consistent with estimates using the JK covariance matrix shown in the table \ref{table:bias}, for all redshift bins. }
\label{tab:mc_cons}
\end{table}

\end{document}